 \documentstyle[pre,aps,epsf,graphicx]{revtex}

\begin{document}


\title {Growth and decay of discrete nonlinear Schr\"odinger breathers 
interacting with internal modes or standing-wave phonons}

\author{Magnus Johansson and Serge Aubry}

\address{Laboratoire L\'eon Brillouin (CEA-CNRS), 
   CEA Saclay, \\
   F-91191 Gif-sur-Yvette Cedex, 
   France \\}


\maketitle

\begin{abstract} 
We investigate the long-time evolution of weakly perturbed single-site 
breathers (localized stationary states) in the discrete nonlinear Schr\"odinger
(DNLS) equation. The perturbations we consider correspond to time-periodic 
solutions of the linearized equations around the breather, and can be either 
(i) 
spatially localized, or (ii) spatially extended. For case (i), which 
corresponds to the excitation of an internal mode of the breather, we find 
that the 
nonlinear interaction between the breather and its internal mode always leads 
to a slow growth of the breather amplitude and frequency. In case (ii), 
corresponding to interaction
between the breather and a standing-wave phonon, the breather will grow 
provided that the wave vector of the phonon is such that the generation of 
radiating 
higher-harmonics at the breather is possible. In other cases, breather decay is
observed. This condition yields a limit value for the breather frequency 
above which no further growth is possible. We also discuss another mechanism 
for 
breather growth and  destruction which becomes important when the amplitude of 
the perturbation is non-negligible, and which originates from the oscillatory 
instabilities of the nonlinear standing-wave phonons.
\end{abstract}





\section{Introduction}


The concept of nonlinear self-localization is of importance for many physical 
phenomena, and has appeared in a number of different contexts since the  
pioneering work by Landau \cite{Landau} on the polaron problem in the 1930s. 
In recent years, much attention has been devoted to studies of spatially 
localized and time-periodic vibrational modes in anharmonic lattices (see e.g. 
\cite{Aubry}, \cite{Flach} for recent reviews).  The 
general existence of such modes, which have been termed {\it discrete 
breathers}, 
or {\it intrinsic localized modes}, as robust solutions to nonlinear (and in 
general non-integrable) lattice-equations was suggested in 1988 by Takeno 
{\it et al } \cite{Takeno}. Later, their existence was rigorously proven under
rather general conditions by MacKay and Aubry \cite{MacKay} by considering the
limit of uncoupled oscillators (the so called {\it anticontinuous} or 
{\it anti-integrable} limit). By means of the implicit function theorem, they 
showed 
that the trivial solution of a single-site localized vibration at the 
uncoupled 
limit could be continued into a localized breather solution for non-zero 
coupling between the oscillators, provided that the individual oscillators are 
anharmonic, and that no multiples of the breather frequency resonate with the 
bands of linear excitations (phonons). As was demonstrated first in 
\cite{Marin96}, the ideas of the rigorous proof can be turned into an 
efficient numerical 
scheme to calculate breather solutions to any desired accuracy. Since the 
discrete breathers appear under very general conditions in anharmonic 
lattices and 
provide efficient means of energy localization, they have been proposed as 
candidates to explain experimentally observed localization of energy in many 
different physical areas, e.g. DNA dynamics \cite{PeyrardDNA}.

Although, from a fundamental and mathematical viewpoint, the existence 
theorems 
for discrete breathers provide an important cornerstone for understanding the 
dynamics of anharmonic lattices, it is probably of even larger physical 
importance to understand the behaviour of a system close to an exact breather 
solution. By linearizing the lattice-equations around the exact solution, one 
can obtain an approximate description of the dynamics of weakly perturbed
 breathers, and in particular the linear stability properties determining
 whether small perturbations will grow exponentially or not. It was shown in 
\cite{Aubry},\cite{MacKay} that the simplest, single-site, breathers are 
generally linearly stable close to the uncoupled limit, and numerical 
investigations using standard Floquet analysis (see e.g. \cite{MarinCrete}) 
have shown 
that linearly stable breathers typically exist also for rather large values 
of the inter-site coupling. However, when considering time-scales large 
compared to the breather period, the mere linear stability of a breather does
no longer guarantee the eternal existence of the breather in the presence of 
small perturbations, and there are still many questions remaining concerning 
the different mechanisms by which breathers may grow or decay, or possibly
finally be destroyed. If the breathers have a finite life-time, the 
determination of this life-time is of large importance for understanding the 
role of 
breathers in real systems.

It is the purpose of this paper to investigate in more detail some mechanisms 
for breather growth and decay in a simple model system, the discrete nonlinear 
Schr\"odinger (DNLS) equation. The DNLS equation is generic in the sense that 
it describes slowly (in time) varying modulational waves in discrete systems 
in 
a 'rotating-wave' approximation (see e.g. \cite{KivsharPeyrard,Daumont}); 
however, due to its extra symmetry properties 
(see Sec.~\ref{sec:perturbation}) it exhibits some 
nongeneric features among discrete systems, e.g. exact quasiperiodic breathers 
\cite{JohanssonAubry}. The single-site breathers of the DNLS equation are 
stationary states which are linearly stable for all inter-site coupling, and
which reduce to the NLS soliton in the continuum limit (see e.g. 
\cite{MacKay,JohanssonAubry,ELS,JohanssonDresden,HennigTsironis,Weinstein}). 
An important 
application appears in nonlinear optics, where the single-site DNLS-breather 
describes a discrete spatial soliton in an array of weakly coupled waveguides 
\cite{waveguides,Aceves}; 
recent experimental observations \cite{Eisenberg} confirm the successful use 
of the DNLS-model in this context. 

Some recent numerical investigations \cite{Kim} have shown that DNLS-breathers 
can be spontaneously created from noisy backgrounds, in a similar manner as 
was previously observed for Klein-Gordon \cite{Daumont,PeyrardKG} and FPU 
\cite{ThierryFPU} lattices. Typically, this spontaneous energy localization 
was observed to occur in two steps. In the first step, a large number of small 
breathers are created as a result of the modulational instability 
\cite{KivsharPeyrard,Daumont,CA,DRT} of travelling plane waves occurring for 
certain wave number regimes. The second step proceeds by inelastic collisions 
between the breathers, in which systematically the big breathers grow at  
expense of the smaller ones. Thus, the outcome will be a small number of large 
breathers, together with some remaining background of small-amplitude (phonon)
oscillations. However, it can generally not be concluded from the numerical 
simulations that this is the true final state of the system, and actually 
long-time simulations for FPU-chains \cite{ThierryFPU} revealed also a third 
step, in which the interaction with the phonon oscillations leads to 
the final destruction of the breather and equipartition of energy. 
Thus, to elucidate the nature of the final states for typical initial 
conditions in anharmonic chains, it is necessary to obtain a better 
understanding of the mechanisms for interactions between breathers and 
small-amplitude perturbations.

In this paper, we take the following approach. As initial state, we consider 
an exact single-site breather solution, and add a small perturbation 
corresponding to a time-periodic eigensolution to the equations of 
motion linearized around the breather. These solutions, which can be either 
localized or extended in space, constitute a complete 
set in which an arbitrary initial perturbation can be expanded. The localized 
solutions
correspond to internal modes of the breather 
\cite{JohanssonDresden,Chen,Baesens,KPCP}, while the excitation of an 
extended solution corresponds to a
standing-wave (i.e., non-propagating) phonon interacting with the breather. 
In Sec.~\ref{sec:perturbation} we describe the model and outline the 
perturbational approach which forms the analytical backbone for the 
interpretation of the numerical results presented in 
Secs.~\ref{sec:internal} and \ref{sec:phonons}. Sec.~\ref{sec:internal} 
discusses the long-time consequences of the interaction between the breather 
and its internal modes, while Sec.~\ref{sec:phonons} concerns the interaction
between the breather and small-amplitude standing-wave phonons of different 
wave vectors. 
We will find that in both cases, a simple argument based on the conservation 
laws can be used to obtain a sufficient condition for breather growth. In 
Sec.~\ref{sec:instabilities} we discuss another type of  mechanism for 
breather growth and 
destruction which becomes appreciable when the amplitude of the standing wave 
is non-negligible (and consequently the perturbational approach can be 
expected to fail), and which has its origin in the recently discovered 
oscillatory instabilities of the nonlinear standing-wave phonon themselves
 \cite{Anna}.
Finally, we make some concluding remarks in Sec.~\ref{sec:concluding}.

Concerning the numerical simulations of the dynamics  presented in this 
paper,  unless otherwise 
stated they always apply for a system of infinite size (finite size systems 
are considered only in Sec.~\ref{sec:instabilities}). The simulations 
have been performed either by using very large system sizes or by appending 
damping regions of various sizes to the boundaries; in all cases we 
have carefully 
checked that the boundary conditions have no essential influence on our 
results.


\section{Model and framework for the perturbational approach}


\label{sec:perturbation}

\subsection{Model}
\label{subsec:model}

We consider the following form of the DNLS Hamiltonian with canonical 
conjugated variables $\{i \psi_n\},\{\psi_n^\ast\}$:
\begin{equation}
\label{DNLSham}
{\cal H} \left( \left\{i \psi_n\right\}, \left\{\psi_n^\ast \right\}\right) = 
\sum_n \left ( C |\psi_{n+1}- \psi_n |^2 - \frac{1}{2}| \psi_n|^4  \right ) 
\equiv \sum_n {\cal H}_n . 
\end{equation}
This yields the DNLS equation
\begin{equation}
\label{DNLS}
i \dot{\psi_n} = \frac {\partial {\cal H }} {\partial \psi_n^\ast } 
= - C(\psi_{n+1} + \psi_{n-1} -2 \psi_n) - | \psi_n|^2 \psi_n ,
\end{equation}
which, in addition to the Hamiltonian (\ref{DNLSham}), also conserves the 
total 
{\em excitation norm} (or {\em power} in nonlinear optics applications),
\begin{equation}
\label{norm}
{\cal N } = \sum_n |\psi_n|^2 \equiv \sum_n {\cal N }_n .
\end{equation}
The conservation laws for the norm and Hamiltonian are, through Noether's 
theorem,  related to the
invariance of the DNLS equation (or, more precisely, of its corresponding 
action integral) under infinitesimal transformations in phase 
($\psi_n \rightarrow \psi_n e ^ {i \epsilon}$) and time 
($t \rightarrow t+\epsilon$), respectively. Defining the 'norm density' 
${\cal N }_n$ and 'Hamiltonian density' ${\cal H}_n$ as in Eqs. (\ref{norm}) 
resp. (\ref{DNLSham}), the conservation laws can be expressed in terms of 
continuity equations as 
\begin{equation}
\label{normcons}
\frac {{\rm d}{\cal N }_n}{{\rm d}t}+\left (J_{\cal N} \right )_{n}
-\left (J_{\cal N} \right )_{n-1} = 0 ,
\end{equation}
\begin{equation}
\label{hamcons}
\frac {{\rm d}{\cal H }_n}{{\rm d}t}+\left (J_{\cal H} \right )_{n}
-\left (J_{\cal H} \right )_{n-1} = 0 ,
\end{equation}
with the {\em (norm) current density}
\begin{equation}
\label{JNdef}
J_{\cal N}  = 2C  {\rm Im} 
\left[\psi_n^\ast \psi_{n+1} \right ] 
\end{equation}
and the {\em Hamiltonian flux density}
\begin{equation}
\label{JHdef}
J_{\cal H} = - 2C {\rm Re} \left[\dot{\psi}_{n+1}(\psi_{n+1}^\ast 
- \psi_{n}^\ast )\right] ,
\end{equation}
respectively. These conservation laws are discrete analogs to those existing 
for the 
continuous NLS equations with general nonlinearities (see e.g. \cite{JJR}), 
however there is no discrete counterpart to the momentum conservation law 
since the discrete equation has no continuous translational symmetry in space. 
Furthermore, we note that the transformation $C\rightarrow -C$ in 
(\ref{DNLS}) is equivalent to 
$\psi_n\rightarrow (-1)^n e^{-i4Ct}\psi_n$, and thus we will for the rest of 
this paper only consider $C>0$ without loss 
of generality.

The single-site DNLS-breather is a stationary-state solution to 
Eq.~(\ref{DNLS}) of the form
\begin{equation}
\psi_n (t)= \phi_n (\Lambda) e^{i \Lambda t} ,
\label{stationary}
\end{equation}
where the time-independent shape $\{\phi_n\}$  depends on the frequency 
$\Lambda$ and is spatially localized with a single 
maximum at a lattice site. The breather exists for all  $\Lambda/C > 0 $; 
the limit $C \rightarrow 0 $ (or $\Lambda \rightarrow \infty $ ) corresponding
to the anticontinuous limit, where $\{\phi_n\}$ is localized at a single 
lattice-site, 
while the limit $\Lambda/C \rightarrow 0 $ corresponds to the continuous 
limit, where $\{\phi_n\}$ approaches the NLS soliton. The single-site breather 
is a ground state solution to Eq.~(\ref{DNLS}) in the sense that it minimizes
the Hamiltonian (\ref{DNLSham}) for a fixed value of the norm (\ref{norm}), 
i.e., $\delta {\cal H} + \Lambda \delta {\cal N} = 0$, 
where the frequency $\Lambda$ appears as the Lagrange multiplier 
(see e.g. \cite{Weinstein}). The norm for the single-site breather, 
${\cal N}_\phi$, is known to be a monotonously increasing 
function of $\Lambda$, while the Hamiltonian, ${\cal H}_\phi$, is negative 
and monotonously decreasing (see e.g.
\cite{Scott,longrange}). From the minimization condition, these functions 
will be related as
\begin{equation}
\frac {{\rm d}{\cal H}_\phi } { {\rm d } \Lambda} =  - \Lambda 
\frac {{\rm d}{\cal N}_\phi } { {\rm d } \Lambda} .
\label{NH}
\end{equation}

To describe the dynamics close to the breather (\ref{stationary}), we 
introduce the following perturbation expansion:
\begin{equation}
\psi_n(t)=\left\{\phi_n+\lambda \epsilon_n ( t ) + \lambda ^2 \eta_n (t) 
+ \lambda^3 \xi_n (t) + \lambda^4 \theta_n (t) +...\right\} 
e ^ {i \int \Lambda dt } ,
\label{perturbation}
\end{equation}
where  $\epsilon_n (0)$ is the initial perturbation and 
$\eta_n(0)=\xi_n(0)=\theta_n(0)= ... = 0$. Thus, as for the usual stability 
analysis of stationary states (see e.g. 
\cite{ELS,Carr}), the perturbation is applied in a frame rotating with the 
breather frequency $\Lambda$. Substituting into Eq.~(\ref{DNLS}) and 
identifying coefficients for consecutive powers of the small parameter 
$\lambda$ yields an infinite set of equations, which from 0th to 4th order 
read:
\begin{eqnarray}
- \Lambda \phi_n + C ( \phi_{n+1}+\phi_{n-1}-2\phi_n)+|\phi_n|^2 \phi_n  
=  0 & &
\label{0th}\\
{\cal L }( \Lambda) \cdot  \{\epsilon_n\}
\equiv
\{i \dot{\epsilon}_n+ C ( \epsilon_{n+1}+\epsilon_{n-1}-2\epsilon_n) 
+ 2 | \phi_n|^2 \epsilon_n + \phi_n^2 \epsilon_n^\ast - \Lambda \epsilon_n \} 
= 0
\label{1st}\\
{{\cal L }( \Lambda) \cdot \{\eta_n\} =  
- \phi_n^\ast \epsilon_n^2 - 2 \phi_n | \epsilon _n |^2}
\label{2nd}\\
{\cal L }( \Lambda) \cdot \{\xi_n\} =  
-2 \phi_n^\ast \epsilon_n \eta_n -2 \phi_n \left( \epsilon_n^\ast \eta_n + 
\epsilon_n \eta_n^\ast \right) - | \epsilon_n|^2 \epsilon_n 
\label{3rd}\\
{\cal L }( \Lambda) \cdot \{\theta_n\} = 
-2 \phi_n \left( \epsilon_n \xi_n^\ast + \epsilon_n^\ast \xi_n + |\eta_n|^2 
\right)
- \phi_n^\ast \left( 2 \epsilon_n \xi_n+\eta_n^2 \right) 
-\epsilon_n^2 \eta_n^\ast -2 |\epsilon_n|^2 \eta_n ,
\label{4th}
\end{eqnarray}
where the operator ${\cal L }( \Lambda)$ (which is linear over the field of 
real numbers) is defined from the first equality in Eq.~(\ref{1st}). The 0th 
order equation (\ref{0th}) gives  the breather 
shape $\{\phi_n\}$ (which for the single-site breather can be assumed real 
and positive without loss of generality), while the 1st order equation 
(\ref{1st}) is the linearization of the DNLS equation around the breather. 

\subsection{Solutions to the linearized equations}

\label{subsec:linear}

To obtain the solutions to the linearized equations (\ref{1st}), we proceed 
in a similar way as is usually done for continuous generalized NLS models 
(see e.g.\cite{Kaup,PKA,ABP}) and introduce a substitution of the form
\begin{equation}
\epsilon_n(t)\equiv\epsilon_n^{(r)}(t)+i \epsilon_n^{(i)}(t) = \frac{1}{2} 
a\left (U_n+W_n \right) e ^{-i \omega_p t } + \frac{1}{2} a ^\ast 
\left (U_n^\ast-W_n^\ast \right) e^{i \omega_p t} ,
\label{subst}
\end{equation}
so that
\begin{equation}
\epsilon_n^{(r)}(t)={\rm Re} \left(\epsilon_n(t)\right) ={\rm Re}\left (a U_n 
e ^{-i \omega_p t } \right) \; , \; \epsilon_n^{(i)}(t)={\rm Im} \left(
\epsilon_n(t)\right) ={\rm Im}\left (a W_n e ^{-i \omega_p t } \right) .
\label{ReIm}
\end{equation}
Substituting  (\ref{subst}) into Eq.~(\ref{1st}) and assuming 
$\phi_n$ real yields 
\begin{eqnarray}
{\cal L }_0 W_n & \equiv  & -C ( W_{n+1} + W_{n-1}-2W_n ) -\phi_n^2 W_n + 
\Lambda W_n  = \omega_p U_n
\label{W_n}\\
{\cal L }_1 U_n & \equiv  & -C ( U_{n+1} + U_{n-1}-2U_n ) - 3 \phi_n^2 U_n + 
\Lambda U_n  = \omega_p W_n,
\label{U_n}
\end{eqnarray}
where the operators ${\cal L }_0$ and ${\cal L }_1$ are Hermitian. 
Thus, we can obtain the eigenfrequencies $\omega_p$ and the corresponding 
eigenvectors $\left( \{U_n\},\{W_n\}\right)$ from matrix diagonalization,
\begin{equation}
\label{matrix}
{\bf M }^{(0)} \cdot  \left ( \begin{array}{c} \{U_n\}\\\{W_n\} \end{array} 
\right) \equiv  \left ( \begin{array}{c} 0 \;\; \; {\cal L}_0 \\ {\cal L}_1 
\;\; \; 
0  \end{array} \right) \left ( \begin{array}{c} \{U_n\}\\\{W_n\} \end{array} 
\right)  = \omega_p \left ( \begin{array}{c} \{U_n\}\\\{W_n\} \end{array} 
\right) .
\end{equation}
To make connection to earlier work \cite{Aubry,MarinCrete}, we remark that 
the vector $\left( \{\epsilon_n^{(r)}\},\{\epsilon_n^{(i)}\}\right)=\left( 
\{U_n\},\{-i W_n\}\right)$ is an eigenvector of the Floquet matrix with 
eigenvalue $e^{-i \omega_p T}$, where the time-period $T$ here is arbitrary 
since the 
operators ${\cal L }_0$ and ${\cal L }_1$ are time-independent. (The 
symplectic Floquet 
matrix is $e^{{\bf M }_{F}T}$, where ${\bf M }_{F}$ is obtained from 
${\bf M }^{(0)}$  by changing ${\cal L }_1$ into $-{\cal L }_1$.) Thus, 
(\ref{subst}) is the linear combination of two complex conjugated 
Floquet eigensolutions which makes $\epsilon_n^{(r)}$ and $\epsilon_n^{(i)}$ 
real. For the single-site breather, all eigenvalues $\omega_p$ of 
${\bf M }^{(0)}$ are always real, implying the linear stability of the 
breather for all parameter values  $\Lambda/C >0$ \cite{Laedke}. Accordingly, 
we can also choose the eigenvectors 
$\left( \{U_n\},\{W_n\}\right)$ of ${\bf M }^{(0)}$ to be real and 
normalized, in which case the phase of the 
amplitude $a$ describes the symmetry properties of the solution (\ref{subst}) 
under time reversal: choosing $a$ real yields a time-symmetric solution, 
$\epsilon_n(t)=\epsilon_n^\ast(-t)$, while choosing $a$ purely imaginary 
yields 
a time-antisymmetric solution, $\epsilon_n(t)=-\epsilon_n^\ast(-t)$.

For an infinite system, the spectrum of the (non-Hermitian) matrix 
$\bf{M}^{(0)}$ can generally 
be divided into a continuous (phonon) part, corresponding to extended 
eigenvectors, and a point spectrum corresponding to localized eigenvectors. 
The phonon 
spectrum for any localized solution $\{\phi_n\}$ is easily obtained from the 
limit $|n|\rightarrow \infty$, since the condition $\phi_n \rightarrow 0$ 
implies that the operators ${\cal L}_0$ and ${\cal L}_1$ become identical 
and  Eqs.~(\ref{W_n}) - (\ref{U_n}) reduce into two uncoupled equations for 
the linear combinations $a_n \equiv U_n+W_n$ resp. $ b_n \equiv U_n-W_n$. 
Assuming $a_n\sim
 e ^ {\pm i q_a n }$ and $b_n\sim e ^ {\pm i q_b n }$, respectively, yields 
the dispersion 
relations
\begin{eqnarray}
\omega_p  & = & \Lambda - 2 C (\cos q_a-1),
\label{adisp}\\
\omega_p  & = & - \Lambda + 2 C( \cos q_b-1),
\label{bdisp}
\end{eqnarray}
from Eqs.~(\ref{W_n}) - (\ref{U_n}). Thus, the continuous spectrum of the 
matrix $\bf{M}^{(0)}$ consists
 of two branches, symmetrically located around $\omega_p=0$, and since 
$\Lambda > 0$ for the single-site breather these two branches never 
overlap. Note also that two eigenvectors with eigenvalues $\pm \omega_p$ 
correspond to the same solution to  Eq.~(\ref{1st}) (changing the sign of 
$\omega_p$ in Eq.~(\ref{subst}) is equivalent to changing $U_n \leftrightarrow 
U_n^\ast$, $W_n \leftrightarrow -W_n^\ast$, $a \leftrightarrow a^\ast$), 
and therefore it is enough to consider e.g. $\omega_p > 0$, in which 
case $b_n = U_n - W_n$ always vanishes exponentially as $n \rightarrow \pm 
\infty$.

\begin{figure}[ht]
\includegraphics[height=16.5cm,angle=270]{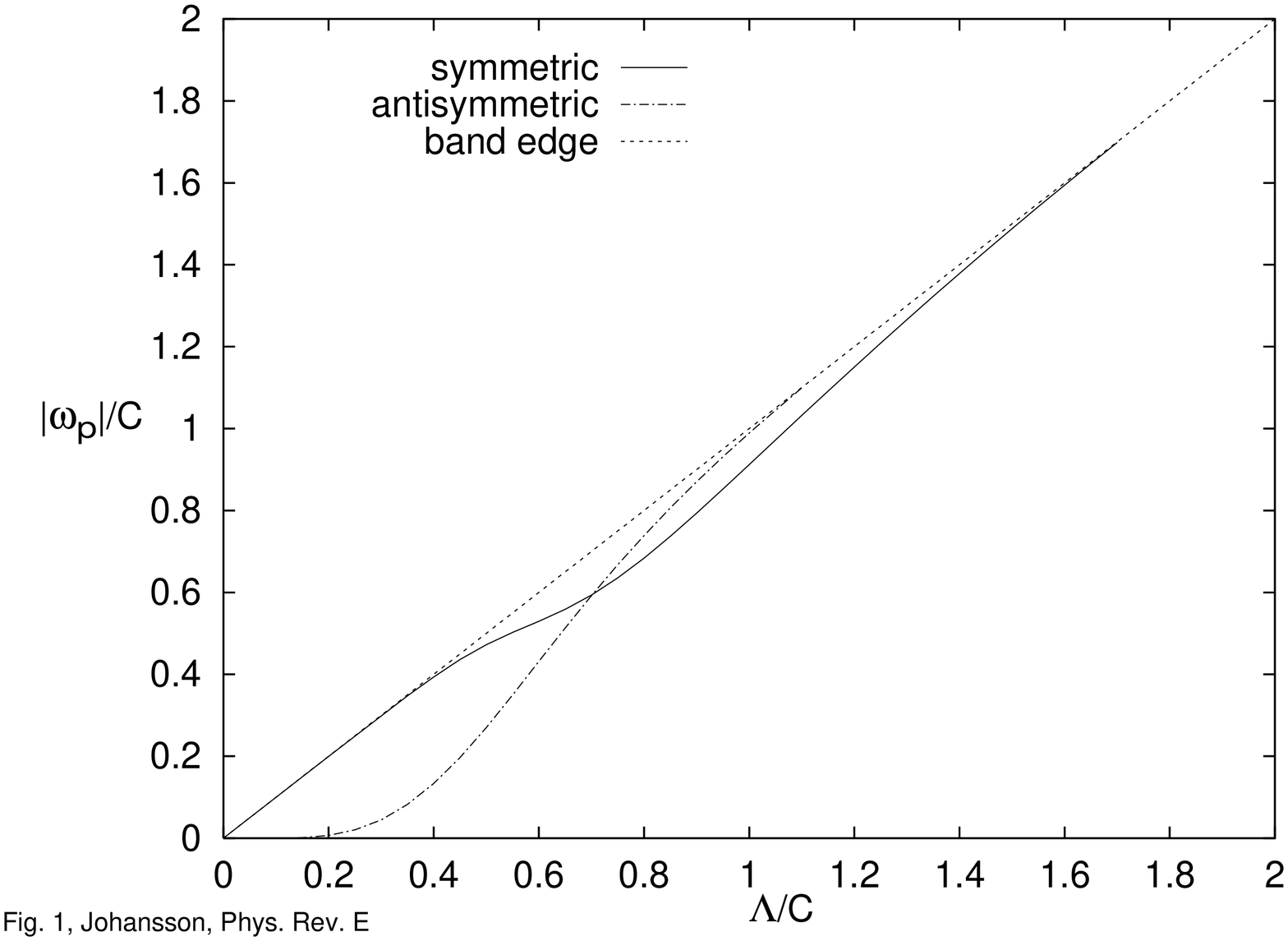}
\vspace{2cm}
\caption{Variation of internal mode frequencies versus breather frequency for
the spatially symmetric (solid line) resp. antisymmetric (dashed-dotted line) 
internal modes of the single-site breather. Dashed straight line shows the 
lower band edge of the phonon band.}
\label{fig1}
\end{figure}

When $\Lambda/C$ is not too large, the linear spectrum around the single-site 
breather contains also two pairs of nonzero isolated eigenvalues $\omega_p$, 
which correspond to the two internal modes of the breather 
\cite{JohanssonDresden,KPCP}. One of these modes is a spatially symmetric, 
'breathing', mode, while the other is a spatially antisymmetric, 
'translational' or 'pinning' mode. Numerically, it has been found that the 
breathing mode 
exists for $0<\Lambda/C \lesssim 1.7$, while the pinning mode exists for 
$0<\Lambda/C \lesssim 1.1$. The numerically calculated internal mode 
frequencies 
$\omega_p$ as a function of breather frequency $\Lambda$ are shown in 
Fig.~\ref{fig1}. Note that as $\Lambda/C\rightarrow 0$, the breathing mode 
frequency 
approaches the lower edge of the phonon band (but always stays outside the band
\cite{KPCP}), while the pinning mode frequency approaches zero (but always
stays nonzero). This is consistent with the fact that the soliton solution of 
the continuous NLS equation has no breathing mode (due to its exact 
integrability), and has a translational mode with zero frequency due to the 
translational symmetry of the NLS equation.

To obtain a complete set of solutions to Eq.~(\ref{1st}) in which an 
arbitrary initial perturbation $\epsilon_n(0)$ can be expanded, we must 
include 
also
the zero-frequency solutions, which generally can be written as a 
superposition of two fundamental modes. One of these modes ('phase mode'
\cite{AubryCretegny}) is the solution $W_n=\phi_n$ to the homogeneous  
equation (\ref{W_n}), ${\cal L}_0 W_n=0$. The corresponding perturbation 
 $\epsilon_n=i\phi_n$ describes 
a rotation of the overall phase of the breather. The second mode 
('growth mode' \cite{AubryCretegny}) is obtained by solving 
the inhomogeneous equation ${\cal L }_1U_n = - \phi_n$ , which has the solution
$U_n=\partial \phi_n / \partial \Lambda$. The corresponding solution to 
Eq.~(\ref{1st}) is 
$\epsilon_n=\partial \phi_n / \partial \Lambda + i \phi_n t $, and 
corresponds to a time-linear growth of 
the perturbation representing a small change in the breather frequency. 

Although the set of eigensolutions (\ref{subst}) together with the 
two zero-frequency modes forms a basis for the space of solutions to 
Eq.~(\ref{1st}) (there are no bifurcations, which could result in additional 
'marginal modes' \cite{AubryCretegny} with time-linear growth at degenerate 
eigenvalues), this basis is in general not orthogonal using the ordinary 
scalar product, and typically there is a 
considerable overlap between the solution corresponding to the internal 
breathing mode
 and the zero-frequency modes. However, in analogy with e.g. 
\cite{Kaup,PKA,ABP}, we can define a 'pseudo-scalar' product between any two 
vectors 
$\left( \{U_n^{(1)}\},\{W_n^{(1)}\}\right)$ and 
$\left( \{U_n^{(2)}\},\{W_n^{(2)}\}\right)$ by
\begin{equation}
\sum_n \left ( U_n^{(1)}W_n^{(2)^\ast}+ W_n^{(1)}U_n^{(2)^\ast} \right) .
\label{scalar}
\end{equation}
This product is formally not a true scalar product,
since for the general case the product of a vector with itself as defined by 
(\ref{scalar}) is not 
necessarily positive. However, when $\left( \{U_n\},\{W_n\}\right)$ is a real 
eigenvector of  $\bf{M}^{(0)}$ we have 
$\sum_n U_n W_n = \frac{1}{\omega_p}\sum_n W_n {\cal L}_0 W_n$ from 
(\ref{W_n}), and the operator ${\cal L}_0$ is positive definite for all $W_n 
\neq \phi_n$ \cite{Laedke}.
 With this product, it follows from (\ref{W_n}) - (\ref{U_n}) that all 
eigensolutions with different (real) eigenfrequencies $\omega_p$ are 
'orthogonal' in the sense that 
\begin{equation}
\left(\omega_p^{(1)}-\omega_p^{(2)}\right)\sum_n \left ( 
U_n^{(1)}W_n^{(2)^\ast}+ W_n^{(1)}U_n^{(2)^\ast} \right) = 0 ,
\label{orthogonal}
\end{equation}
 and 
the only nonzero product involving the zero-frequency modes is the 
cross-product between the phase mode and the growth mode \cite{Laedke}:
\begin{equation}
\sum_n \phi_n \frac {\partial \phi_n } { \partial \Lambda } = \frac {1} {2} 
\frac {{\rm d}{\cal N}_\phi } { {\rm d } \Lambda} > 0 ,
\label{overlap}
\end{equation}
where ${\cal N}_\phi$ is the norm (\ref{norm}) of the breather with frequency 
$\Lambda$. We also remark that, since the product (\ref{scalar}) 
multiplied by a factor $i$ is just the symplectic product between the two 
vectors $\left( \{\epsilon_n^{(r)(1)}\},\{\epsilon_n^{(i)(1)}\}\right)$ and 
$\left( \{\epsilon_n^{(r)(2)}\},\{\epsilon_n^{(i)(2)}\}\right)$, the sign of 
the product of an eigenvector with itself can 
be interpreted as the negative of the Krein signature of the corresponding 
pair of Floquet eigenvalues \cite{Aubry}.

\subsection{Strategy for the perturbational approach}

\label{subsec:strategy}

We now consider as initial state a single-site breather perturbed in the 
direction of a single eigenmode (\ref{subst}) (localized or extended) of the 
linearized equations (\ref{1st}), and wish to describe qualitatively the 
long-time effects of this perturbation using the expansion 
(\ref{perturbation}). In general, taking into account terms up to order $p$ 
in this expansion yields a solution to the DNLS equation which is correct to 
$O(\lambda^{p+1})$, i.e., for long but finite time-scales for small 
initial perturbations (note that the expansion parameter $\lambda$ plays the 
same role as the mode amplitude $a$.).
As is wellknown however, this kind of expansion in general 
diverges due to resonances between solutions to the homogeneous equation 
(\ref{1st}) and the inhomogeneous terms appearing in the right-hand sides of 
Eqs. (\ref{2nd})-(\ref{4th}) and the corresponding higher-order equations. A 
resonance with a solution belonging to the continuous spectrum results in a 
bounded but nonlocalized solution corresponding to outgoing radiation, 
while a resonance with an eigenfunction 
belonging to the discrete spectrum gives a spatially localized response which 
diverges linearly in 
time. However, up to any finite order these divergences can be systematically 
removed by 
allowing a slow time dependence of the independent variables, which in our 
case are taken to be the mode amplitude $a$ and the
 breather frequency $\Lambda$. This procedure  adds additional terms to the 
equations, which 
can be tuned so that the divergent parts of the response disappear.
 In other words, these two quantities are used as 
collective variables which, together with the outgoing radiation fields, are 
expected to describe the main features of the asymptotic dynamics if the 
initial perturbation is sufficiently small.

The second order correction is given 
by the inhomogeneous Eq.~(\ref{2nd}), which with the substitution  
(\ref{subst}) becomes (choosing $\phi_n$, $U_n$, and $W_n$ real without loss 
of generality):
\begin{eqnarray}
\label{2nd2}
{\cal L }( \Lambda) \cdot \{\eta_n\}  =  - \frac {1} {2} \phi_n \left[|a|^2
\left( 3 U_n^2 + W_n^2 \right)
 + \left( 3 U_n^2-W_n^2 \right) {\rm Re}\left (a^2e ^{-2 i \omega_p t } 
\right)\right.\nonumber \\
\left.+2 i U_n W_n {\rm Im}\left (a^2e ^{-2 i \omega_p t }\right)\right] .
\end{eqnarray}
Thus, the right-hand side contains one static part and 
one part involving the frequencies $\pm 2 \omega_p$. It 
acts as a periodic force with frequencies 0 and $2 \omega_p$, and since all 
terms contain the factor $\phi_n$ this force is localized at the breather 
region. The response to this force will remain bounded and localized unless 
the corresponding homogeneous equation (\ref{1st}) has a solution with 
frequency 0 resp. $ 2 \omega_p$ which is non-orthogonal to the 
corresponding part of the right-hand side in (\ref{2nd2}). As will be shown 
in Sec.~\ref{subsec:higher-order}, a non-zero overlap between 
the static part of (\ref{2nd2}) and the zero-frequency solutions of 
(\ref{1st}) yields a (time-independent) shift of the breather frequency. 
Moreover, if $\Lambda<2|\omega_p|<\Lambda+4C$, 
so that $2 \omega_p$ is inside the phonon band of the homogeneous equation, 
a resonance will generally occur, resulting in radiation with frequency $2 
\omega_p$ 
emitted from the breather region. The strength of the radiation field is 
determined by the (generally nonzero) overlap between the $ 2 \omega_p$-part 
of (\ref{2nd2}) and 
the corresponding homogeneous solution (see Sec.~\ref{subsec:higher-order}). 
In a 
similar way, we obtain that the 
right-hand-side of the third order equation (\ref{3rd}) contains the 
frequencies $\omega_p$ and $3\omega_p$, the fourth order equation (\ref{4th}) 
contains the 
frequencies 0, $2\omega_p$, and $4\omega_p$, and in general the {\it p}th 
order equation contains as its highest harmonic the frequency $p\omega_p$.
Accordingly, we conclude that if 
\begin{equation}
\Lambda<p|\omega_p|<\Lambda+4C,
\label{pradiation}
\end{equation}
 so 
that $p\omega_p$ belongs to the phonon band, the perturbed breather will 
radiate to {\it p}th order. The consequences of this radiation for the 
breather itself will be discussed in Secs.~\ref{sec:internal} and 
\ref{sec:phonons} for the cases of localized and extended perturbations 
$\{\epsilon_n\}$, respectively.


\section{Breather interacting with internal modes}


\label{sec:internal}

With the initial perturbation $\epsilon_n(0)$ of the single-site breather 
corresponding to a spatially localized eigenmode of the linearized equations
(\ref{1st}), it is clear from the discussion in Sec.~\ref{subsec:strategy} 
that higher order radiation 
always will be created, since for any internal mode frequency $\omega_p$ 
there is always an integer $p$ such that $p\omega_p$ belongs to the phonon 
band and (\ref{pradiation}) is fulfilled. Moreover, from the numerical 
results presented in Fig.~\ref{fig1} we find that the spatially symmetric 
breathing mode always radiates to second order, since (\ref{pradiation}) 
always is fulfilled for $p=2$, while the antisymmetric pinning mode radiates 
to second order only when $\Lambda/C \gtrsim 0.480$. Thus, due to this 
radiation from the breather the total norm contained in any finite region 
around the breather(i.e., the total norm of breather + internal mode) will 
always decrease with time. However, the main concern here is the long-time 
effect of the internal-mode excitation on the breather itself, and thus we 
must investigate whether there will also be some transfer of energy between 
the breather and its internal mode. We will first (Sec. \ref{subsec:numerics})
show results from direct numerical integration of Eq. (\ref{DNLS}); then we 
will give two alternative approaches to the analytical 
interpretation of these results based on the higher-order equations 
(\ref{2nd})-(\ref{4th}) (Sec. \ref{subsec:higher-order}) resp. 
the conservation laws (\ref{normcons})-(\ref{hamcons}) 
(Sec. \ref{subsec:conservation}).

\subsection{Numerical simulations}

\label{subsec:numerics}

\begin{figure}[ht]
\noindent
\includegraphics[height=8.8cm,angle=270]{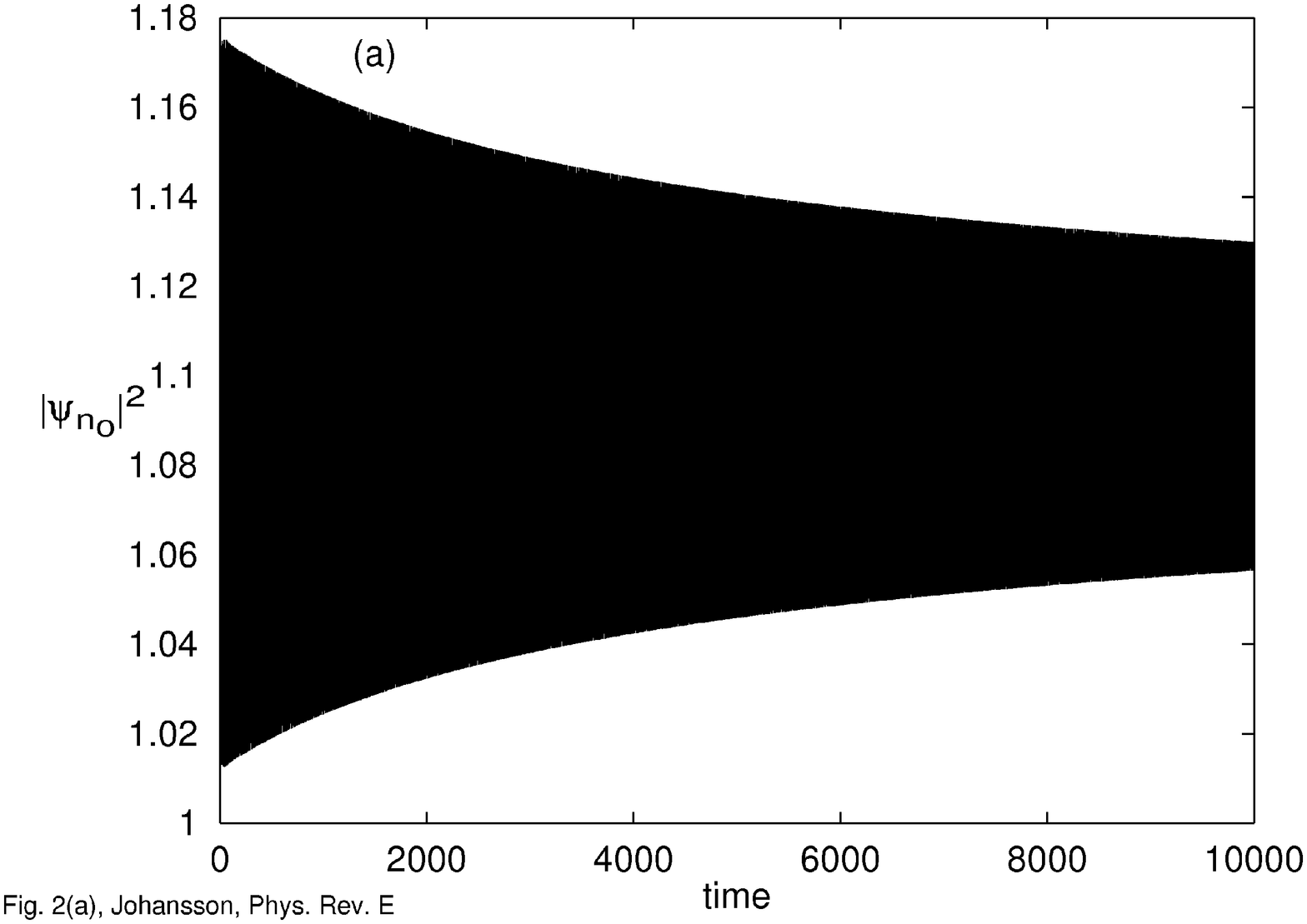}
\includegraphics[height=8.8cm,angle=270]{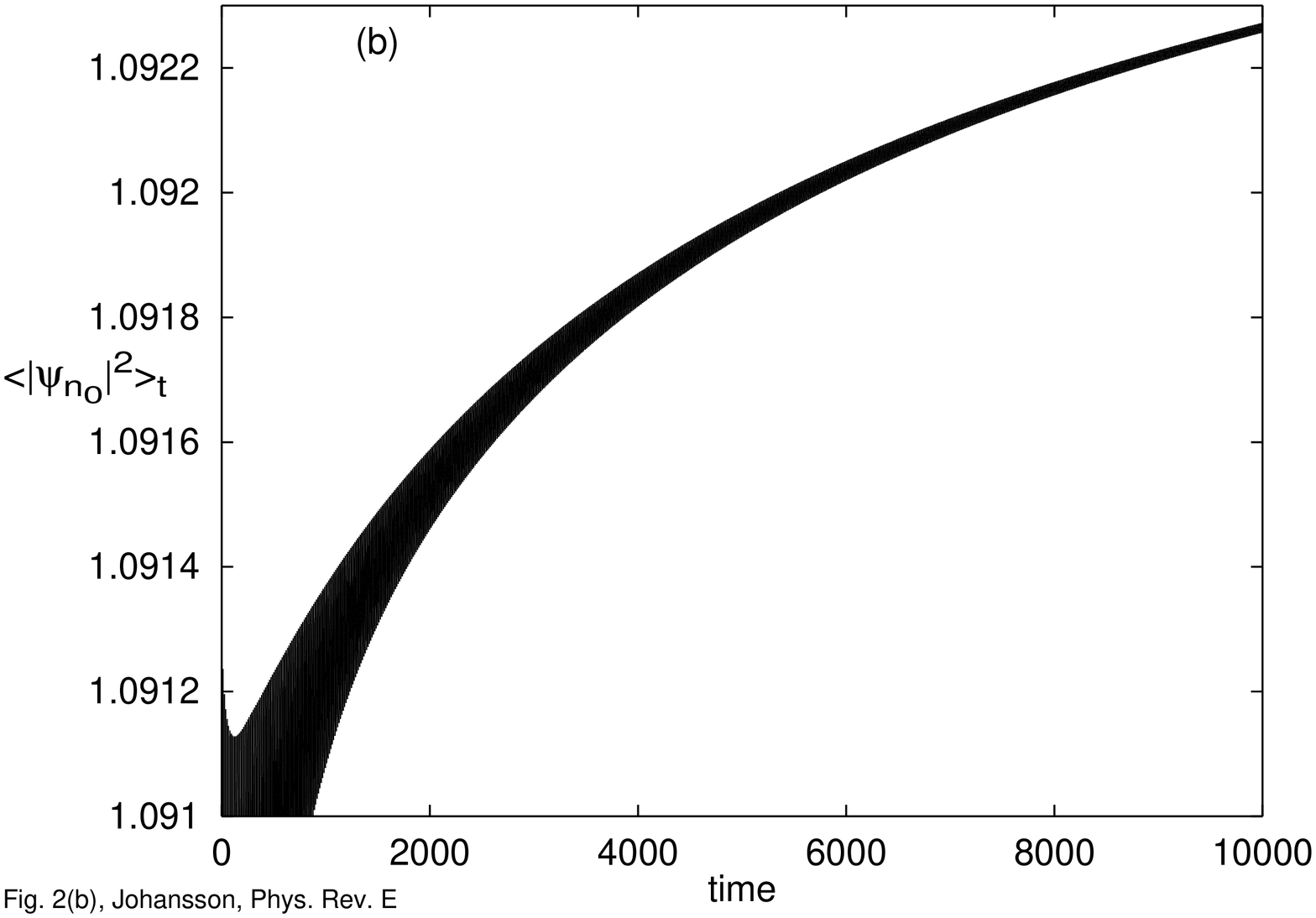}
\includegraphics[height=8.8cm,angle=270]{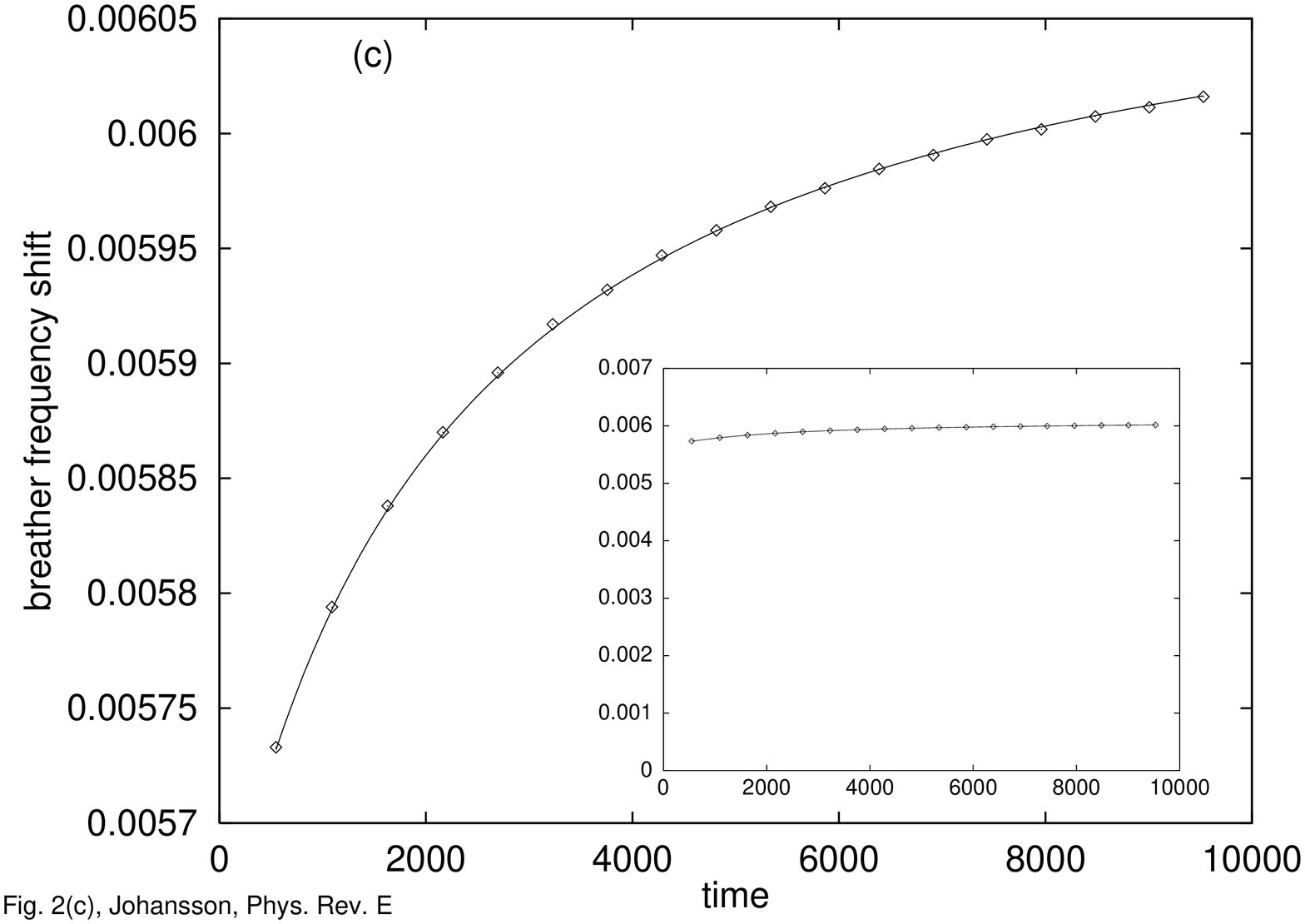}
\vspace{0.5cm}
\caption{Time-evolution of a breather with an initial perturbation in the 
direction of the breathing mode. Parameters are $\Lambda=0.5$, 
$\omega_p \approx 0.47$, and $C=1$. (a) shows the time-evolution of the 
central-site intensity $|\psi_{n_0}|^2$, (b) shows its time-average 
$\left \langle |\psi_{n_0}|^2 \right \rangle_t$ calculated using 
Eq.~(\ref{average}), while (c) shows the instantaneous shift of breather 
frequency $\Lambda(t)-\Lambda_0$. The solid line in the main figure in (c) is 
a fit using Eq.~(\ref{shift}) with $a(0)=0.082$, $C_1=0.9078$, $C_2=0.069$, 
and $\gamma=0.067$.}
\label{fig2}
\end{figure}

In Fig.~\ref{fig2}, we show a typical example on the long-time evolution of a 
breather 
when the initial perturbation is taken in the direction of its internal 
breathing mode. As is seen from Fig.~\ref{fig2} (a), the amplitude of the 
breathing mode decays slowly with time as a consequence of the losses due to 
generation of second order radiation, and a careful study of its 
envelope $|a(t)|$ indicates that it decays as 
\begin{equation}
|a(t)| \sim \frac{|a (0)|} {\sqrt{1+\gamma |a(0)|^2 \, t}} ,
\label{decay}
\end{equation}
 where $\gamma > 0 $ is a constant. This is consistent with a similar 
result obtained for the continuum NLS equation with generalized (non-cubic) 
nonlinearity \cite{PKA}; the analytical motivation for this result (which is 
analogous to that of the continuum model given in Ref. \cite{PKA}) is given 
in the following subsections.

However, the main result of this section is 
illustrated in Fig.~\ref{fig2} (b) and (c). Fig.~\ref{fig2} (b) is obtained by 
calculating the time-average of the central-site intensity as
\begin{equation}
\left \langle|\psi_{n_0}|^2 \right \rangle_{t=t_K}=\frac{1}{K}\sum_{k=1}^K|
\psi_{n_0}(t_k)|^2,
\label{average}
\end{equation}
where $t_k$ is a set of closely spaced time-instants. It is clear that the 
interaction between the breather and its internal mode asymptotically leads to 
an {\it increase} of the average peak intensity, i.e., to 
{\it breather growth}. The same phenomenon is illustrated also in 
Fig.~\ref{fig2} (c), where we have plotted the difference between the 
instantaneous breather 
frequency calculated at time $t$, $\Lambda(t)$, and the frequency of the 
unperturbed breather 
$\Lambda_0$. From this figure, we can also conclude that there are two 
different mechanisms causing the shift of breather frequency. Firstly, there 
is an initial (almost instantaneous) rather large frequency shift, which can 
be interpreted as an adaption of the initially perturbed breather to the 
breather which is 'closest' to the initial condition. As is shown below 
in Sec. \ref{subsec:higher-order}, this time-independent frequency shift, 
which is 
observed to be always positive for the breathing mode, is a consequence 
of 
the overlap between the static part of the right-hand side of the second-order 
equation (\ref{2nd2})  and the zero-frequency modes.
Secondly, there is the slow, continuous increase of the 
breather frequency which corresponds to the slow increase of 
$\left \langle |\psi_{n_0}|^2 \right \rangle_t$ in Fig.~\ref{fig2} (b), 
indicating a continuous transfer of norm from 
the internal mode to the breather. It is described by the static part of the 
right-hand side of the fourth-order equation (\ref{4th}) (see Sec. 
\ref{subsec:higher-order}).

\begin{figure}[ht]
\noindent
\includegraphics[height=13.8cm,angle=270]{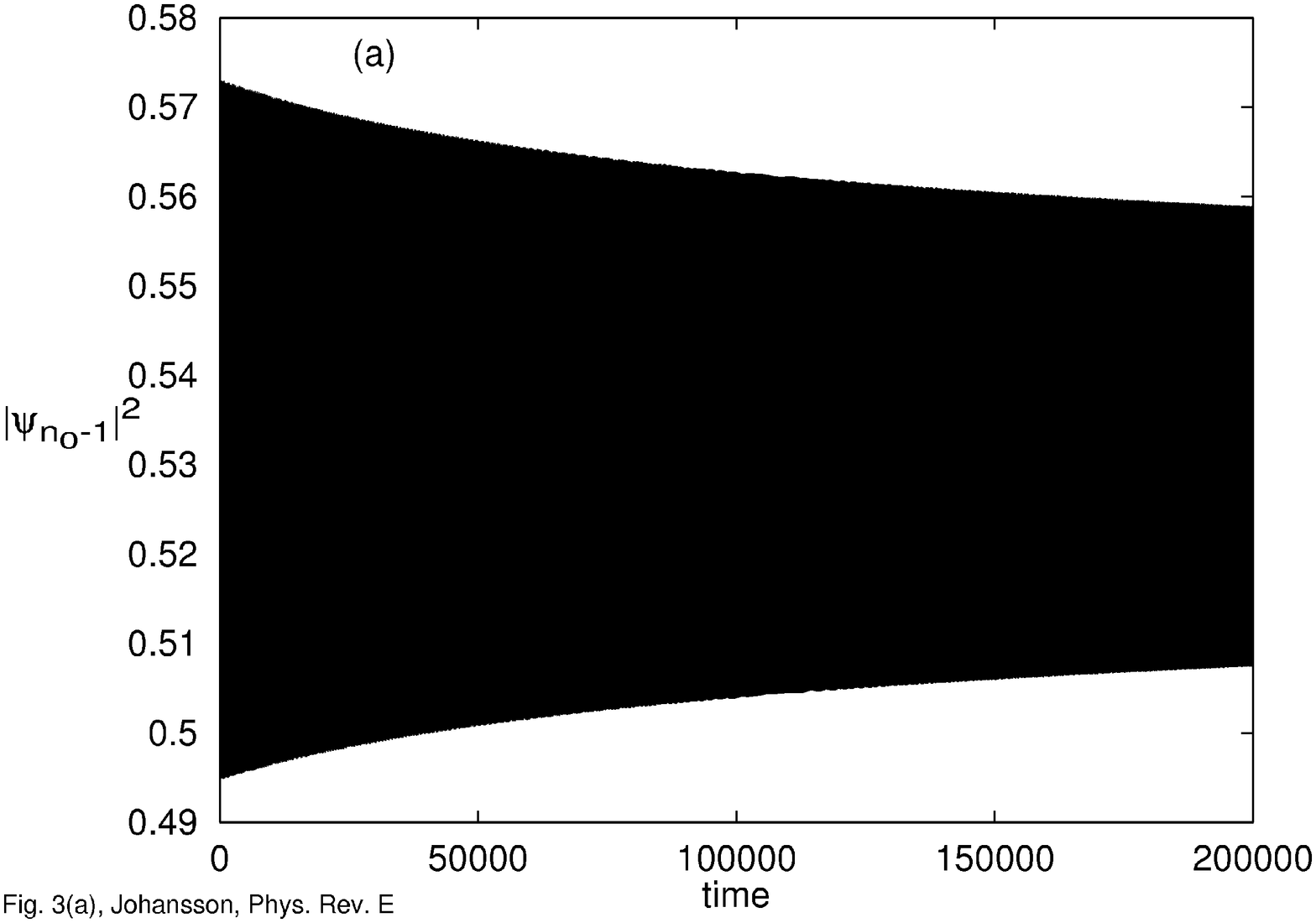}
\includegraphics[height=13.8cm,angle=270]{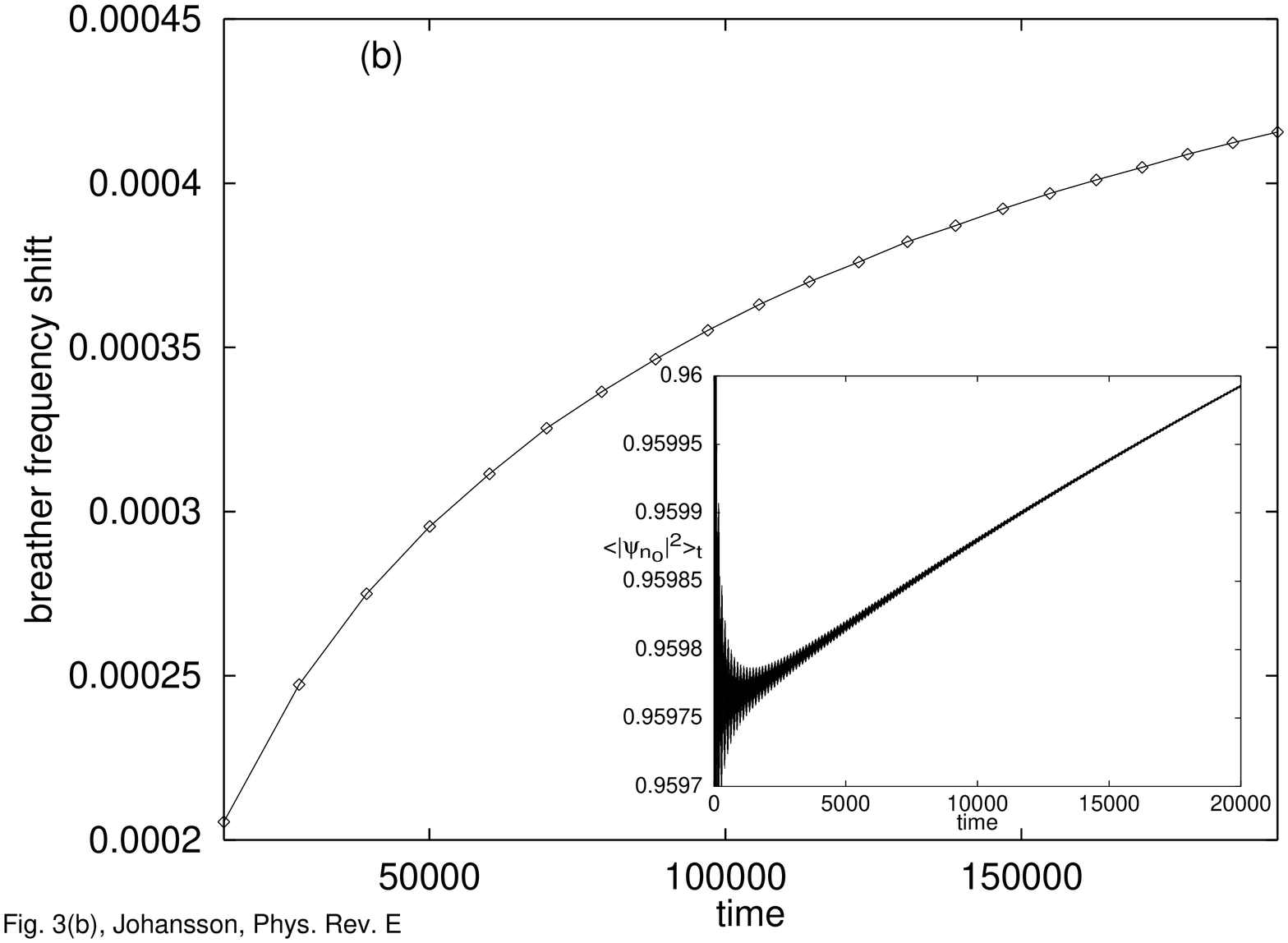}
\vspace{0.5cm}
\caption{Time-evolution of a breather with an initial perturbation in the 
direction of the pinning mode. Parameters are $\Lambda=0.45$, 
$\omega_p \approx 0.197$, and $C=1$. (a) shows the time-evolution of 
$|\psi_{n_0-1}|^2$, where $n_0$ is the central site of the breather, 
(b) (main figure) shows the instantaneous shift of breather 
frequency $\Lambda(t)-\Lambda_0$, and inset in (b) shows the time-average 
$\left \langle |\psi_{n_0}|^2 \right \rangle_t$.} 
\label{fig3}
\end{figure}

When the initial perturbation of the breather is taken in the direction of 
its internal 
pinning mode we observe, just as for the breathing mode, that the 
breather-internal mode interaction asymptotically {\it always leads to 
breather 
growth}. An example is shown in Fig.~\ref{fig3}, where the parameter values 
have been chosen so that the lowest harmonic that enters 
the phonon band is $3 \omega_p$ ($\Lambda/C=0.45<0.480$), We observe two 
qualitative differences compared to the case with breathing mode excitation. 
Firstly, since in this case the first phonon resonance occurs only in the 
third order equation (\ref{3rd}), the decay of the internal mode amplitude 
will be slower, and a good fit 
is obtained by $|a(t)| \sim |a(0)|\left (1+\gamma |a(0)|^4 \, t \right )
^{-1/4}$. This agrees with the  general result 
when $p \omega_p$ is the lowest harmonic that enters the phonon band obtained 
for the continuum generalized NLS equation in Ref. \cite{PKA}; the derivation 
of the corresponding result for the discrete case (see Eq.~(\ref{ageneral})) 
is given in Sec.~\ref{subsec:conservation}.
As a consequence of the slower decay of the internal mode amplitude, 
the breather growth will also be slower when $p > 2$, as can be seen from 
Fig.~\ref{fig3} (c) by comparing the time-scales with those of 
Fig.~\ref{fig2}. 
Secondly, the initial 
shift of the breather frequency for a pinning mode excitation is always much 
smaller than for the breathing mode 
excitation (also when $2 \omega_p$ is in the phonon band), and is when 
$\Lambda/C \gtrsim 0.55$ also observed to be negative. The explanation for this 
is given in Sec. \ref{subsec:higher-order}. However, it is important to 
stress that also in the cases where the initial frequency shift is negative,  
we find that
the 
continuous breather growth  always will give an asymptotic frequency shift 
which is positive.

\subsection{Analysis of higher-order equations}

\label{subsec:higher-order}

Here, we will analyze the higher-order equations (\ref{2nd})-(\ref{4th}) by 
making 
use of the strategy of systematically removing the appearing divergent parts 
as outlined in Sec. \ref{subsec:strategy} (in analogy with the treatment of 
the continuous NLS-type equations in e.g. Refs. \cite{Kaup,PKA,ABP}). First, 
we show how the dominating contribution to the time-independent frequency 
shift observed in the numerical simulations above can be calculated from the 
static part of the right-hand side of the second-order equation (\ref{2nd}).
This frequency shift can be explicitly taken into account
by replacing $\Lambda$ in Eq.~(\ref{perturbation}) with 
$\Lambda_0+\lambda^2 \Lambda_2$, where $\Lambda_0$ is the unperturbed 
breather frequency and $\Lambda_2$ the second-order shift to be determined.
This implies that the additional term $\Lambda_2 \phi_n$ will be added to the 
right-hand side 
of Eq.~(\ref{2nd2}). 
Writing the response to the static part of (\ref{2nd2}) as 
$\eta_n^{(s)} = |a|^2 ( u_n^{(s)} + i  w_n^{(s)}) $ with real $u_n^{(s)}$ and  
$w_n^{(s)}$ then yields
\begin{equation}
\label{Mstatic}
{\bf M }^{(0)} \cdot \left ( \begin{array}{c} \left\{u_n^{(s)}\right\}\\
\left\{w_n^{(s)}\right\} \end{array} \right)=\left ( \begin{array}{c} \left\{
0\right\}\\\left\{-\phi_n \left(\frac{\Lambda_2}{|a|^2}-\frac{1}{2} \left(
3U_n^2+W_n^2\right)\right)\right\} \end{array} \right) ,
\end{equation}
with ${\bf M }^{(0)}$ as defined by Eq. (\ref{matrix}). If the expansion of 
the 
right-hand side of (\ref{Mstatic}) in the complete set of vectors consisting 
of the eigenvectors of 
${\bf M }^{(0)}$ (including the phase mode) and the growth mode contains some 
component on either of the two zero-frequency 
modes, the response $\eta_n^{(s)}$ will not remain bounded but diverge 
linearly with time. Thus,
in order to remove this divergency, the frequency shift
$\Lambda_2$ must be chosen so that both these components are identically zero. 
The component on the growth mode is trivially zero, while the 
component on the phase mode is obtained by applying the pseudo-scalar product 
(\ref{scalar}) with 
the vector corresponding to the growth mode and using (\ref{overlap}). 
Demanding this component to be zero yields
\begin{equation}
\Lambda_2=\frac{|a|^2}{\frac {{\rm d}{\cal N}_\phi } { {\rm d } \Lambda}} 
\sum_n \phi_n \frac {\partial \phi_n } { \partial \Lambda } 
\left(3U_n^2+W_n^2\right).
\label{static}
\end{equation}
This is typically positive for the breathing mode, since the dominating 
contribution to the sum in (\ref{static}) comes from the central site $n_0$, 
and 
$\frac {\partial \phi_{n_0} } { \partial \Lambda }$ is always positive. For 
the spatially antisymmetric pinning mode, there is no contribution to this 
sum from the central site, since
$U_{n_0}$ and $W_{n_0}$ are zero. The 
change from a positive to a negative frequency shift when increasing 
$\Lambda$ in this case is related to a qualitative change 
of the nature of the growth mode $\frac {\partial \phi_{n} } { \partial 
\Lambda}$: for $\Lambda\gtrsim 0.55$ we find that $\frac {\partial \phi_{n} } 
{ \partial \Lambda}<0$ for all $n\neq n_0$, so that all terms in the sum in 
Eq.~(\ref{static}) are 
negative, while $\frac {\partial \phi_{n} } { \partial \Lambda}$ becomes 
positive also for sites in the neighborhood of $n_0$ for smaller $\Lambda$.

For the rest of the analysis in this subsection, we assume for calculational 
simplicity that 
the internal mode frequency is such that $2\omega_p$ is inside the phonon band 
(and thus it is not applicable for the pinning mode excitation when 
$\Lambda/C \lesssim 0.480$). 
Then, the non-static part of the right-hand side of the second order equation 
(\ref{2nd2}) will generally give rise to a non-localized response, which can 
be written in the form 
$\eta_n^{(rad)}= \frac{1}{2}a^2 \left( u_n^{(2)}+w_n^{(2)} \right) e^{-2 i 
\omega_p t} +\frac{1}{2}a^{\ast 2 } \left( u_n^{(2)^\ast}-w_n^{(2)^\ast} 
\right) e^{2 i \omega_p t}$. This response corresponds to the radiation field 
going out 
from the breather region, and since the right-hand side of (\ref{2nd2}) 
is spatially localized and symmetric, this field should asymptotically 
correspond to two identical linear waves propagating to the left (right) for
$n \rightarrow - \infty$ ($ + \infty$). Thus, the boundary 
conditions can be written as
\begin{equation}
\label{BC}
 u_n^{(2)},w_n^{(2)} \rightarrow r_2 e ^ {\pm i q_2 n } \; , \; n \rightarrow 
\pm \infty , 
\end{equation}
with $q_2 = \arccos \left ( 1 - \frac {2\omega_p-\Lambda}{2C} \right)$ 
according 
to (\ref{adisp}). 
Defining for general 
$\omega$ the matrix $\bf{M}^{(\omega)}$ (cf. (\ref{matrix})) as
\begin{equation}
\label{Momega}
{\bf M }^{(\omega)} \equiv  \left ( \begin{array}{l} -\omega \;\; \;\;
{\cal L}_0 \\\; {\cal L}_1 \;\; -\omega  \end{array} \right) ,
\end{equation}
the functions $u_n^{(2)}$ and $w_n^{(2)}$ are seen from (\ref{2nd2}) to be 
determined by
\begin{equation}
\label{M2omegap}
{\bf M }^{(2\omega_p)} \cdot  \left ( \begin{array}{c} \{u_n^{(2)}\}\\\{
w_n^{(2)}\} \end{array} \right)  = \frac{\phi_n}{2} \left ( \begin{array}{c} 
\{2U_n W_n\}\\\{3 U_n^2-W_n^2\} \end{array} \right) .
\end{equation}
Since for general $\omega$ every eigenvector of ${\bf M }^{(\omega)}$ 
with eigenvalue $\mu$ is also an eigenvector of ${\bf M }^{(0)}$ with 
eigenvalue $\mu+\omega$, the right-hand side can be expanded on the basis of 
eigenvectors of ${\bf M }^{(0)}$ (including the zero-frequency modes). The 
strength of the radiation field is then 
given by the expansion coefficient for the (continuous spectrum) eigenvector 
of ${\bf M }^{(0)}$ 
with 
eigenvalue $2 \omega_p$, since this corresponds to the eigenvalue $\mu=0$ of 
${\bf M }^{(2\omega_p)}$, and thus a spatially non-bounded response in 
(\ref{M2omegap}).
Using the orthogonality relation (\ref{orthogonal}), this coefficient is 
simply 
the 'overlap' between the right-hand side of (\ref{M2omegap}) and the 
eigenvector of ${\bf M }^{(0)}$ with eigenvalue $2 \omega_p$ calculated with 
 the pseudo-scalar product (\ref{scalar}), which is 
generally nonzero.

Next, we show how the dominating contribution to the decay  of the 
internal mode amplitude as given by Eq.~(\ref{decay}) is obtained from the 
condition that 
$\xi_n$ in the 
third-order equation (\ref{3rd}) should remain bounded. To this end, we
 assume a 
slow time-dependence of the internal mode amplitude of the form 
$a=a(\lambda ^2 t)$, and consider the response to 
the terms with frequency $\omega_p$ in the right-hand side of Eq.~(\ref{3rd}) 
'corrected' by the additional terms 
$(-i \dot{a} + \Lambda_2 a ) (U_n+W_n) e^{- i \omega_p t} + (-i \dot{a}^\ast 
+ \Lambda_2 a^\ast ) (U_n-W_n) e^{ i \omega_p t}$ appearing as a consequence 
of 
including the time-dependence of $a$ and the second-order frequency shift 
$\Lambda_2$ from Eq.~(\ref{static}) in the perturbation expansion 
(\ref{perturbation}).
Writing the response to this part as $\xi_n^{(\omega_p)}= \frac{1}{2} |a|^2 a 
\left( u_n^{(3)}+w_n^{(3)} \right) e^{- i \omega_p t} +\frac{1}{2} |a|^2 
a^\ast \left( u_n^{(3)^\ast}-w_n^{(3)^\ast} \right) e^{ i \omega_p t}$ yields
\begin{eqnarray}
\label{Momegap}
\lefteqn
{{\bf M }^{(\omega_p)} \cdot \left ( \begin{array}{c} \{u_n^{(3)}\}\\
\{w_n^{(3)}\} \end{array} \right)  } 
\nonumber \\
&
=
&
\left ( \begin{array}{c} \{i \frac{\dot{a}}{|a|^2 a} U_n + \phi_n  \left(U_n 
w_n^{(2)}-W_n u_n^{(2)}+ 2 W_n u_n^{(s)}\right)+ \frac{1}{4}\left(U_n^2 W_n
+3W_n^3 \right) - \frac{\Lambda_2} {|a|^2} W_n\}\\\{i \frac{\dot{a}}{|a|^2 a} 
W_n + \phi_n  \left(3U_n u_n^{(2)}+W_n w_n^{(2)}+6U_n u_n^{(s)}\right)+ 
\frac{1}{4}\left(3U_n^3+U_nW_n^2 \right) - \frac{\Lambda_2} {|a|^2} U_n\} 
\end{array} \right) .
\end{eqnarray}
A bounded response for $\xi_n^{(\omega_p)}$  exists
only if the vector on the right-hand side of (\ref{Momegap}) has no component  
in the direction of the internal mode eigenvector 
$\left( \{U_n\},\{W_n\}\right)$, since this is the eigenvector corresponding 
to the eigenvalue zero of ${\bf M }^{(\omega_p)}$. Using the orthogonality 
relation  (\ref{orthogonal}), this component is obtained by application of 
the product (\ref{scalar}) with the vector $\left( \{U_n\},\{W_n\}\right)$, 
and the condition that this component must be zero determines the 
time-evolution of $a$. 
Considering only the absolute value $|a|^2$, the resulting 
equation has the form $ \frac {\rm d}{{\rm d}t} \left ( |a|^2 \right)+ \gamma 
|a|^4 = 0 $, which has the desired solution (\ref{decay}). The constant 
$\gamma$ is given by
\begin{equation}
\label{gamma}
\gamma = \frac{\sum_n \phi_n  \left [ 2 U_n W_n {\rm Im} \left ( w_n ^{(2)}
\right) + \left(3 U_n^2 - W_n ^2\right)  {\rm Im} \left ( u_n ^{(2)}\right)
\right] } {\sum_n U_n W_n } = \frac{8 C \omega_p |r_2|^2 \sin q_2 } 
{\sum_n W_n {\cal L}_0 W_n } > 0 ,
\end{equation}
where the second equality is obtained using Eqs.~(\ref{W_n}), (\ref{BC}) and 
(\ref{M2omegap}), 
and the positivity of $\gamma$ follows from the fact that, as mentioned in 
Sec.~\ref{subsec:linear}, 
the operator  ${\cal L}_0$ is positive definite for all $W_n \neq \phi_n$ 
\cite{Laedke}.

Finally, we show how the continuous increase of the breather frequency 
appears from the divergent response to the static part of the 
right-hand side of the fourth-order equation (\ref{4th}). In its unmodified 
form, this part is given by 
\begin{eqnarray}
R_n^{(4s)} \equiv - \frac{1}{2} \phi_n |a|^4 \left\{6U_n {\rm Re} 
\left(u_n^{(3)}\right)+2 W_n {\rm Re} \left(w_n^{(3)}\right )
+2 i U_n {\rm Im} \left(w_n^{(3)}\right) -2iW_n {\rm Im} \left(u_n^{(3)}\right)
\right.
\nonumber \\
\left.
+6 (u_n^{(s)})^2 +3 |u_n^{(2)}|^2
+|w_n^{(2)}|^2+2i{\rm Im} \left(u_n^{(2)\ast}w_n^{(2)}\right)
\right\}  
\nonumber\\
- \frac{1}{4}|a|^4\left\{
2\left(3U_n^2+W_n^2\right)u_n^{(s)}
+\left(3U_n^2-W_n^2\right){\rm Re} \left(u_n^{(2)}\right)
+ 2U_nW_n{\rm Re} \left(w_n^{(2)}\right) 
\right. 
\nonumber \\
\left.
-2i U_n W_n {\rm Im} \left(u_n^{(2)}\right)
+ i \left(U_n^2-3W_n^2 \right)
{\rm Im} \left(w_n^{(2)}\right)\right\} .%
\label{static4}
\end{eqnarray}
Now, it is clear from (\ref{static}) and (\ref{decay}) that the 
time-dependence of $a$ will induce a time-dependence of the second-order 
frequency shift $\Lambda_2$ of the form $\Lambda_2(\lambda^2 t)$, so that we 
can express the total breather frequency up to order $\lambda^4$ as
$\Lambda(t)=\Lambda_0+\lambda^2\Lambda_2(\lambda^2 t)+\lambda^4\Lambda_4$, 
where a 4th order correction $\Lambda_4$ has also been included. Then, we must 
also take into account the time-dependence of the breather shape $\phi_n$ by 
writing $\phi_n(\Lambda(t))$. As a consequence, 
the term
$-i \frac {\partial \phi_{n} } { \partial \Lambda} \dot{\Lambda}_2+ \Lambda_4 
\phi_n$
will be added to the expression (\ref{static4}) for $R_n^{(4s)}$ in the 
right-hand side of (\ref{4th}), and  
writing the response to this total force as 
$\theta_n^{(s)} = |a|^4 \left(u_n^{(4s)} + i  w_n^{(4s)} \right) $ with real 
$u_n^{(4s)}$ and  
$w_n^{(4s)}$ yields
\begin{equation}
\label{M4}
{\bf M }^{(0)} \cdot \left ( \begin{array}{c} \{u_n^{(4s)}\}\\\{w_n^{(4s)}\} 
\end{array} \right)
=\frac {1} {|a|^4}\left ( \begin{array}{c} \left\{\frac {\partial \phi_{n} } 
{ \partial \Lambda}\dot{\Lambda}_2 - {\rm Im} \left(R_n^{(4s)}\right)\right\}
\\
\left\{- \Lambda_4 \phi_n 
-{\rm Re} \left(R_n^{(4s)}\right)\right\} \end{array} \right) .
\end{equation}
The response $\theta_n^{(s)}$ will be bounded in time only if the right-hand 
side of (\ref{M4}) has no component either on the growth mode or on the phase 
mode, which gives two conditions for the determination of $\dot{\Lambda}_2$ 
and $\Lambda_4$. Using (\ref{scalar})-(\ref{overlap}), we obtain by demanding 
the expansion coefficient for the growth mode to be zero:
\begin{eqnarray}
\label{Lambda2dot}
\dot{\Lambda}_2=\frac{|a|^4}{2\frac {{\rm d}{\cal N}_\phi }
 { {\rm d } \Lambda}} 
\sum_n \left\{4\phi_n^2\left[ W_n{\rm Im}\left (u_n^{(3)}\right)
-U_n {\rm Im}\left (w_n^{(3)}\right)
+{\rm Im}\left (u_n^{(2)}w_n^{(2)\ast}\right)\right]
\right.
\nonumber\\
\left.
+\phi_n \left[2 U_n W_n 
{\rm Im}\left (u_n^{(2)}\right)+\left(3W_n^2-U_n^2\right) 
{\rm Im}\left (w_n^{(2)}\right) \right] \right\} .
\end{eqnarray}
(Similarly, $\Lambda_4$ is obtained by demanding the 
component on the phase mode to be zero.) Thus, the dominating contribution to 
the frequency growth 
should be of order 
$\dot{\Lambda}_2\sim |a|^4$, so that with the approximate time-dependence 
(\ref{decay}) of the internal-mode amplitude we obtain qualitatively
\begin{equation}
\Lambda(t)-\Lambda_0 \sim |a(0)|^2 \left(C_1 - C_2 \frac {1} {1+
\gamma |a(0)|^2 \, t } \right ) , 
\label{shift}
\end{equation}
which is in good agreement with the numerically observed time-dependence of 
the frequency shift as 
shown in Fig.~\ref{fig2} (c). However, the positivity of $\dot{\Lambda}_2$ 
is not easily seen from the expression (\ref{Lambda2dot}), and therefore we 
will in the following subsection derive an alternative expression from which 
the positivity follows immediately, using  the conservation laws for the norm 
resp. Hamiltonian.

\subsection{Approach using conservation laws}

\label{subsec:conservation}

We consider first the conservation law (\ref{normcons}) for the total norm 
(\ref{norm}) contained in any large but finite region around the breather. 
Averaging over a time-interval $[t,t+2\pi/\omega_p]$ and using Eqs. 
(\ref{perturbation}), (\ref{subst}), we can write the time-averaged norm to 
second order in $|a|$ (putting $\lambda=1$) as
\begin{equation}
\langle {\cal N } \rangle _t (t) \simeq \sum_n \left[\phi^2_n(\Lambda(t))+ 
\frac{|a(t)|^2}{2}\left(U_n^2+W_n^2\right)\right] .
\label{normav}
\end{equation}
(Note that there will be no contribution at order 2 from the static 
second-order 
correction $u_n^{(s)}$, since the renormalization of the breather frequency 
$\Lambda$ according to (\ref{static}) yields $\sum_n \phi_n u_n^{(s)} =0$.)
In the case when $2\omega_p$ belongs to the phonon band, we obtain  the 
following balance equation, which is correct up to order $|a|^4$:
\begin{equation}
\frac{{\rm d}\langle {\cal N } \rangle _t}{{\rm d}t} 
 = \frac {{\rm d}{\cal N}_\phi } { {\rm d } \Lambda} \dot{\Lambda} 
+\frac{1}{2}\sum_n \left(U_n^2+W_n^2\right) \frac {{\rm d} |a|^2}{{\rm d}t} 
= J_{\cal N}(-\infty)- J_{\cal N}(+\infty) = -4C|a|^4|r_2|^2\sin q_2 , 
\label{dNdt}
\end{equation}
where we have used Eqs.~(\ref{normcons}), (\ref{JNdef}) and (\ref{BC}), and 
$q_2$ is as defined below Eq.~(\ref{BC}). 

Similarly, we can use the conservation law (\ref{hamcons}) for the 
Hamiltonian (\ref{DNLSham}) together with the general expression for the 
Hamiltonian flux density (\ref{JHdef}) for a small-amplitude plane wave 
$\psi_n=A e^{i(q n - \omega (q) t)}$, 
\begin{equation}
J_{\cal H} = 2 |A|^2 C \omega(q) \sin q, 
\label{hamflowpw}
\end{equation}
to
write the balance equation for the total time-averaged Hamiltonian in the 
same region for the case of second order radiation:
\begin{equation}
\frac{{\rm d}\langle {\cal H } \rangle _t}{{\rm d}t}  = 
\frac {{\rm d}{\cal H}_\phi } { {\rm d } \Lambda} \dot{\Lambda} 
+ \frac {\partial \langle{\cal H}\rangle_t } { \partial |a|^2 }\frac {{\rm d} 
|a|^2}{{\rm d}t} 
= J_{\cal H}(-\infty)- J_{\cal H}(+\infty) = -4C|a|^4|r_2|^2(2\omega_p
- \Lambda) \sin q_2 , 
\label{dHdt}
\end{equation}
which is also correct to order $|a|^4$. The lowest order contribution to the 
derivative 
$\frac {\partial \langle{\cal H}\rangle_t } { \partial |a|^2}$ can be obtained 
using the first equality in the 
equation of motion (\ref{DNLS}) and its complex conjugate as follows:
\begin{eqnarray}
\nonumber
\frac {\partial \langle{\cal H}\rangle_t } { \partial |a|^2}  
= \frac{1} {a^\ast}\frac {\partial \langle{\cal H}\rangle_t } { \partial a}
=\frac {1} {a^\ast} \sum_n  \left \langle \frac { \partial { \cal H }}
{\partial \psi_n} \cdot \frac {\partial \psi_n} { \partial a} 
+ \frac { \partial  { \cal H }}{\partial \psi_n^\ast} \cdot 
\frac {\partial \psi_n^\ast} { \partial a} \right \rangle_t  
\\
= -\frac{1}{2} \Lambda \sum_n \left(U_n^2+W_n^2\right) 
+ \omega_p \sum_n U_nW_n + {\cal O} (|a|^2),
\label {dHda2}
\end{eqnarray}
which is always negative for an internal mode excitation since  
$|\omega_p| < \Lambda$. Thus, we can combine the two balance equations 
(\ref{dNdt}) and (\ref{dHdt}), and using Eqs.~(\ref{NH}) and (\ref{dHda2}) 
we obtain the expression (\ref{decay}) for 
$|a(t)|$ with $\gamma$ as in (\ref{gamma}), together with the following 
expression for the frequency growth rate from which its positivity is 
immediately seen:
\begin{equation}
\dot{\Lambda}  = \frac { 4C|a|^4|r_2|^2\sin q_2 }{\frac{{\rm d}{\cal N}_\phi } 
{ {\rm d } \Lambda} }\left(\frac{ \sum_n \left(U_n^2+W_n^2\right)}
{ \sum_n U_nW_n} -1 \right) > 0 .
\label{Lambdadot}
\end{equation}

This approach also has the advantage that it is easily generalized to the case 
where the lowest harmonic that enters the phonon band is $p\omega_p$ with 
$p>2$, i.e., for the pinning mode excitation when $\Lambda/C\lesssim 0.480$.
Then, we can write the boundary conditions at the infinities to lowest order 
in $a$ as 
$ \psi_n \rightarrow a^p r_p e ^ {i [\pm q_p n -(p \omega_p - \Lambda)t]}$, 
$n \rightarrow \pm \infty $, where $r_p \sim 1$ and 
$q_p = \arccos \left ( \frac {\Lambda-p\omega_p}{2C} + 1 \right)$ according 
to (\ref{adisp}). Consequently, we can proceed exactly as above, writing
down the balance equations for the norm and Hamiltonian to order $|a|^{2p}$ 
just by modifying the right-hand sides of Eqs.~(\ref{dNdt}) resp. (\ref{dHdt}) 
by replacing $|a|^4$ with  $|a|^{2p}$, $r_2$ with $r_p$, $q_2$ with $q_p$, and 
$2\omega_p$ with $p\omega_p$. Combining the balance equations yields the 
following general expression for the time-dependence of the internal mode 
amplitude,
\begin{equation}
|a(t)| = \frac{|a(0)|}{ \left [1+(p-1)\gamma_p |a(0)|^{2p-2}t \right]^
{1/(2p-2)}}\;,\; \gamma_p = \frac{4pC|r_p|^2 \sin q_p } {\sum_n U_n W_n } > 0 ,
\label{ageneral}
\end{equation}
which is the analog to the expression obtained with similar arguments in 
\cite{PKA} for the continuum NLS models. And most importantly, we obtain a 
general 
expression for the breather frequency growth rate which is positive for all 
$p$:
\begin{equation}
\dot{\Lambda}  = \frac {4C|a|^{2p}|r_p|^2\sin q_p}{\frac{{\rm d}{\cal N}_\phi }
 { {\rm d } \Lambda} }\left(\frac{ p \sum_n \left(U_n^2+W_n^2\right)}
{ 2 \sum_n U_nW_n} -1 \right) > 0 .
\label{Lambdadotgeneral}
\end{equation}
Thus, integrating Eq.~(\ref{Lambdadotgeneral}) using the time-dependence 
(\ref{ageneral}) of the internal-mode amplitude, we obtain that the 
dominating contribution to the frequency growth generally can be written
qualitatively as 
\begin{equation}
\Lambda(t)-\Lambda_0 \sim |a(0)|^2 \left(C_1 - C_2 \frac {1} {1+(p-1)\gamma_p 
|a(0)|^{2p-2} \, t } \right ) ^{1/(p-1)}.
\label{shiftgeneral}
\end{equation}

Let us finally point out that the approach used in this section provides a 
simple physical interpretation of why the interaction between the 
breather and its internal mode in addition to generate radiation also should 
lead to breather growth. Since the expression (\ref{hamflowpw}) for the 
Hamiltonian flux density of a small-amplitude plane wave
is positive when $\omega(q)$ and $q$ have the same sign, the 
Hamiltonian energy for a plane wave always propagates in the same direction 
as the wave itself. Thus, the second (or higher) order radiation emitted from 
the 
breather region will always carry 
away a {\em positive} amount of the Hamiltonian energy, or, equivalently, 
negative 
Hamiltonian energy will flow into the breather region. Moreover, from 
Eq.~(\ref{dHda2}) it is clear that the
contribution to the Hamiltonian from the internal mode always is negative and 
a monotonously {\em decreasing} function of its amplitude, and thus 
the decay of the internal mode would cause an {\em increase} of the 
Hamiltonian in the breather region. Consequently, since the 
Hamiltonian of the pure breather is a monotonously decreasing function of the 
breather frequency, 
the breather should grow so that the total Hamiltonian of (breather+internal 
mode) decreases. A similar mechanism was recently found 
to cause soliton growth in the parametrically driven continuum NLS equation in 
the regime of oscillatory instability \cite{ABP}, and this type of argument 
has also been used to explain the 'quasi-collapse' of a broad excitation to 
a narrow localized state in the two-dimensional DNLS equation \cite{CGMMRRRT}.


\section{Breather interacting with standing-wave phonons}
\label{sec:phonons}


We choose, as in the previous section, the initial perturbation 
$\epsilon_n(0)$ to be an eigensolution of the linearized equations 
(\ref{1st}), but now we consider the case of a spatially extended 
perturbation. Choosing without loss of generality a solution of the form 
(\ref{subst}) with positive 
frequency $\omega_p$ yields the asymptotic behaviour
\begin{equation}
U_n,W_n
\rightarrow \cos (q n \pm \delta ) ; \ n \rightarrow \pm \infty ,
\label{standingwave}
\end{equation}
where the wave vector $q$ $(0 \leq q \leq \pi ) $ is determined by the 
dispersion relation (\ref{adisp}), and $\delta$ is the phase shift across the 
breather. Thus, the excitation of an extended eigenmode corresponds to an
interaction between the breather and a non-propagating, standing-wave phonon 
with small amplitude $a$. As was mentioned in the introduction, these 
standing-wave phonons are generally unstable, but since the instabilities
become exponentially weak in the small-amplitude limit they are expected to 
 have very 
little effect on the breather for the perturbation sizes and time-scales 
considered in this section. We will return to the effects on the breather of
these instabilities in Sec.~\ref{sec:instabilities}, where larger 
perturbations are considered.

In contrast to the case of excitation of a localized internal mode discussed 
in 
Sec.~\ref{sec:internal}, where higher order radiation always was emitted 
from the breather region, the condition (\ref{pradiation}) yields that 
for the standing-wave perturbation the breather will radiate to higher order 
only if $\Lambda \leq \omega_p < \Lambda/2 + 2C $, so that both $\omega_p$ and 
$ 2\omega_p$ are inside the phonon band. In terms of the phonon wave vector 
$q$, this 
means that there is a critical value $q_c$,
\begin{equation}
q_c = \arccos \left( \frac{\Lambda}{4C} \right) ,
\label{qc}
\end{equation}
such that for $0 \leq q < q_c$ second order radiation will be emitted from 
the breather region, while for $q_c<q \leq \pi$ ($\Lambda/2+2C< \omega_p \leq 
\Lambda+4C$) {\em all} multiples of $\omega_p$ are outside the phonon band, 
and no higher order radiation is emitted. As will be shown below, these two 
regions yield qualitatively different scenarios for the long-time evolution 
of the perturbed breather. We also note that for $\Lambda > 4C$ we have 
$q_c=0$, so that for the highly localized, high-frequency breathers {\em no 
phonons 
can generate higher-order radiation} (note also that there are no internal 
modes in this regime). Furthermore, we always have $q_c < \pi / 2 $, so that 
the regime of higher-order radiation generation is a subset of the regime 
$0<q<\pi/2$ where modulational instability for {\em travelling} plane waves 
occurs \cite{KivsharPeyrard}.

\begin{figure}[ht]
\noindent
\includegraphics[height=13.2cm,angle=270]{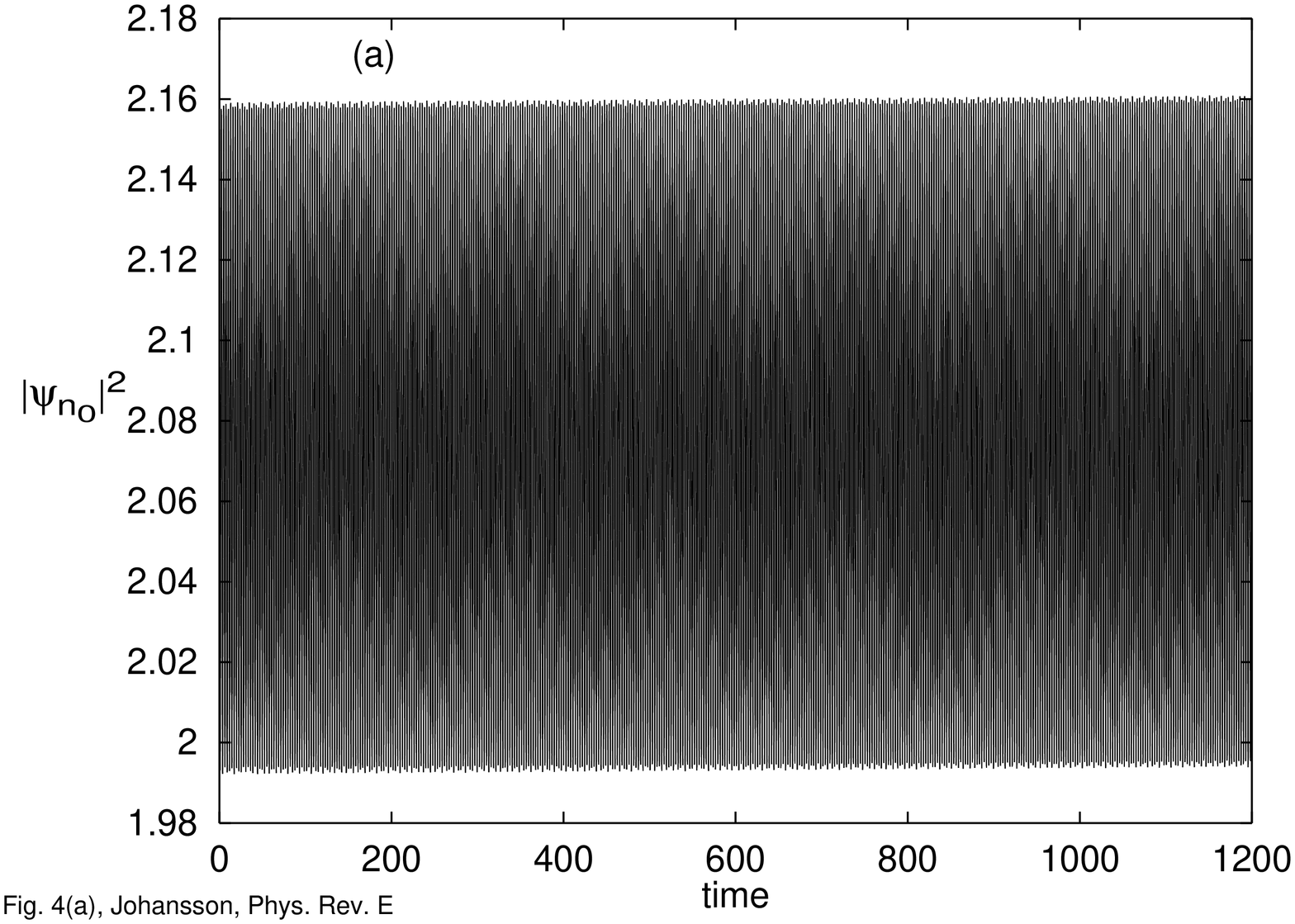}
\includegraphics[height=13.2cm,angle=270]{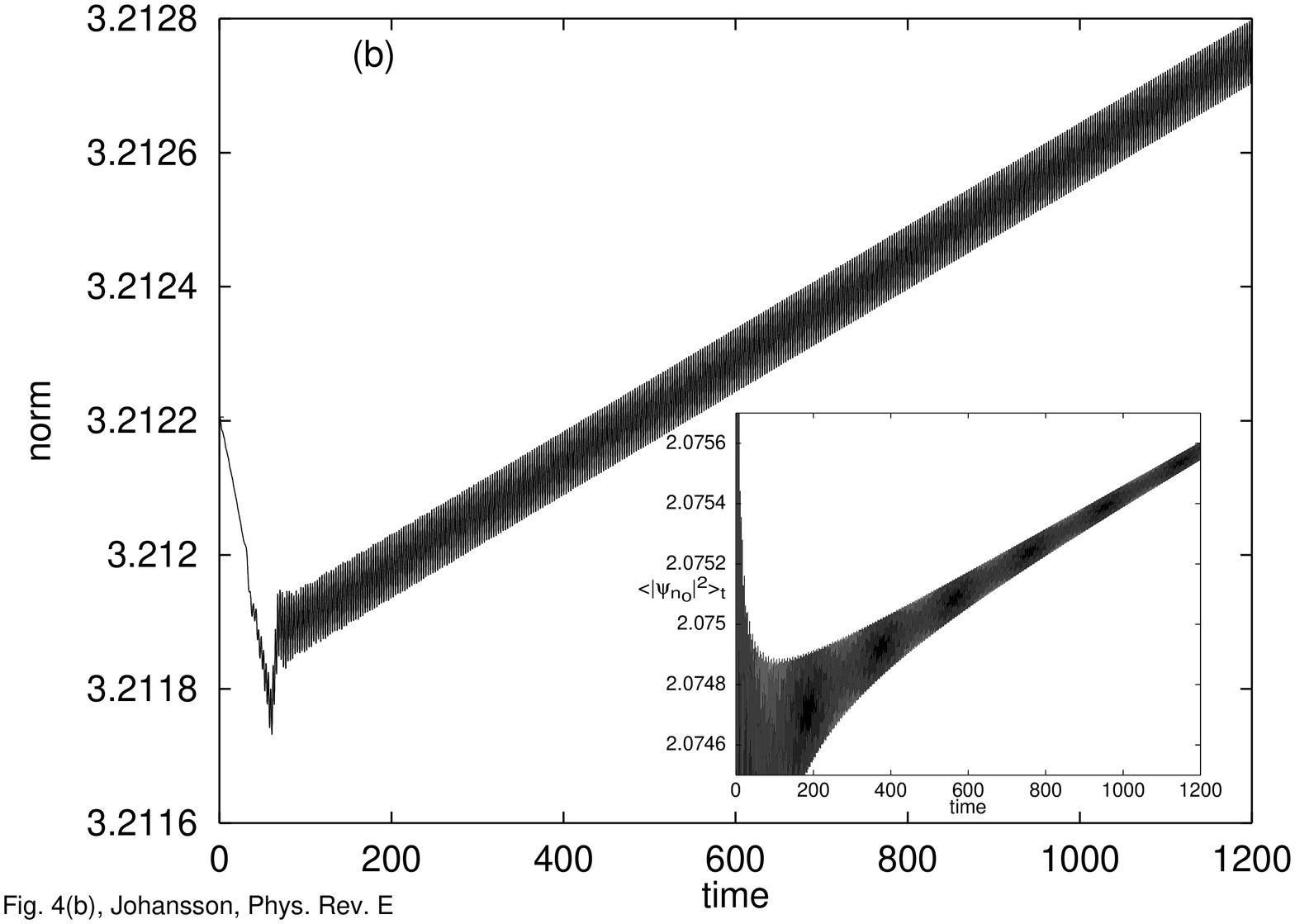}
\vspace{0.5cm}
\caption{Time-evolution of a breather with an initial perturbation in the 
direction of an extended eigenmode (spatially symmetric) with $q<q_c$. 
Parameters are $\Lambda=1.0$, 
$\omega_p \approx 2.31$ ($q \approx 1.22$), $a \approx 0.0383$, and $C=1$. (a) 
 shows the 
time-evolution of the central-site intensity $|\psi_{n_0}|^2$, 
(b) (main figure) shows the total norm contained in a region of 120 sites 
around the 
breather, and inset in (b) shows the time-average 
$\left \langle |\psi_{n_0}|^2\right \rangle_t$.} 
\label{fig4}
\end{figure}

Let us first discuss the case $q<q_c$. A typical example of the long-time 
evolution of a breather interacting with a 
small-amplitude standing-wave phonon with $q<q_c$ is illustrated by 
Fig.~\ref{fig4}. As is seen from Fig.~\ref{fig4} (a), the amplitude of the 
oscillations remains essentially constant in time, but a closer inspection 
reveals that the average value of $|\psi_{n_0}|^2$ asymptotically 
{\em increases} with an apparently constant rate (see inset 
in Fig.~\ref{fig4} (b)). Similarly, Fig.~\ref{fig4} (b) shows that also the 
total norm contained in any finite region around the breather asymptotically 
increases linearly with time. We find that these results are generic for all 
cases when the 
phonon wave vector $q<q_c$ (the spatial symmetry of the phonon is not 
important for the asymptotic behaviour), and thus we conclude that in this 
regime, the breather can 'pump' energy from the phonon (which is infinite for 
an infinite system), and thereby grow. 

In the same spirit as for the internal mode excitation in 
Sec.~\ref{subsec:conservation}, we can give a simple argument 
based on the conservation laws to motivate why the generation of second order 
radiation should lead to 
breather growth. To this end, we assume that the initial standing-wave phonon 
is 
infinitely extended, and that far away from the breather a stationary regime 
will be reached corresponding to the following boundary conditions
\begin{equation}
\psi_n \rightarrow \left[ (a e ^ {\mp i q n } + r e ^{\pm iq n })
e^ {-i \omega_p t } + 
r_2 e ^ { i (\pm q_2 n - 2 \omega_p t) } \right] e ^{i\Lambda t} 
\; ; \; n \rightarrow \pm \infty .
\label{scattering}
\end{equation}
Thus, we have taken into account the second-order radiation with frequency 
$2 \omega_p$ generated at the 
breather region but neglected possible higher-order radiation; moreover the 
resonance at the original phonon frequency $\omega_p$ in the third-order 
equation (\ref{3rd}) has been taken into account by allowing the incoming and 
outgoing complex amplitudes $a$ and $r$ to be different. We can then, in 
analogy with Eqs.~(\ref{dNdt}) and (\ref{dHdt}), write the balance equations 
for the total norm and Hamiltonian contained in a region around the breather 
averaged over a time-interval 
$[t,t+2\pi/\omega_p]$ as
\begin{equation}
\frac{{\rm d}\langle {\cal N } \rangle _t}{{\rm d}t} 
 = \frac {{\rm d}{\cal N}_\phi } { {\rm d } \Lambda} \dot{\Lambda} 
=\langle J_{\cal N}(-\infty)\rangle_t- \langle J_{\cal N}(+\infty)\rangle_t 
= 4C\left[(|a|^2-|r|^2)\sin q -|r_2|^2\sin q_2 \right], 
\label{dNdtph}
\end{equation}
and
\begin{eqnarray}
\nonumber
\frac{{\rm d}\langle {\cal H } \rangle _t}{{\rm d}t}  = 
\frac {{\rm d}{\cal H}_\phi } { {\rm d } \Lambda} \dot{\Lambda} 
&=&
\langle J_{\cal H}(-\infty)\rangle_t - \langle J_{\cal H}(+\infty)\rangle_t  
\\
&=& 4C\left[(|a|^2-|r|^2|)
(\omega_p-\Lambda)\sin q -|r_2|^2(2\omega_p- \Lambda) \sin q_2\right] , 
\label{dHdtph}
\end{eqnarray}
respectively. Here, we have used the facts that the time-average of the 
norm current density (\ref{JNdef}) and the Hamiltonian flux density 
(\ref{JHdef}) are additive quantities for small-amplitude plane waves, 
and that in the stationary regime, the mode amplitudes $a$, 
$r$ and $r_2$ are time-independent. Combining Eqs.~(\ref{dNdtph}) and 
(\ref{dHdtph}) and using (\ref{NH}), we obtain that the breather frequency 
grows with a constant rate given by
\begin{equation}
\dot{\Lambda}  = \frac { 4C|r_2|^2\sin q_2 }{\frac{{\rm d}{\cal N}_\phi }
 { {\rm d } \Lambda} }  > 0 .
\label{Lambdadotph}
\end{equation}
The physical interpretation of this result is, similarly as for the case of 
internal mode excitation, that the generation of higher-order radiation 
results in a net flow of negative Hamiltonian energy into the breather region, 
which is absorbed by the breather by increasing its frequency and maximum 
amplitude. This process is similar to the one observed for the two-channel 
phonon scattering on breathers in a Klein-Gordon model with a Morse potential 
in Ref. \cite{CAF}; however in the latter case the second outgoing wave 
resulted from a resonance in the linearized equations and were therefore of 
the same order of magnitude as the incoming wave, and moreover the outcome 
in this case was breather decay since the energy of a Klein-Gordon breather is 
an increasing function of its amplitude.

\begin{figure}[ht]
\noindent
\includegraphics[height=8.0cm,angle=270]{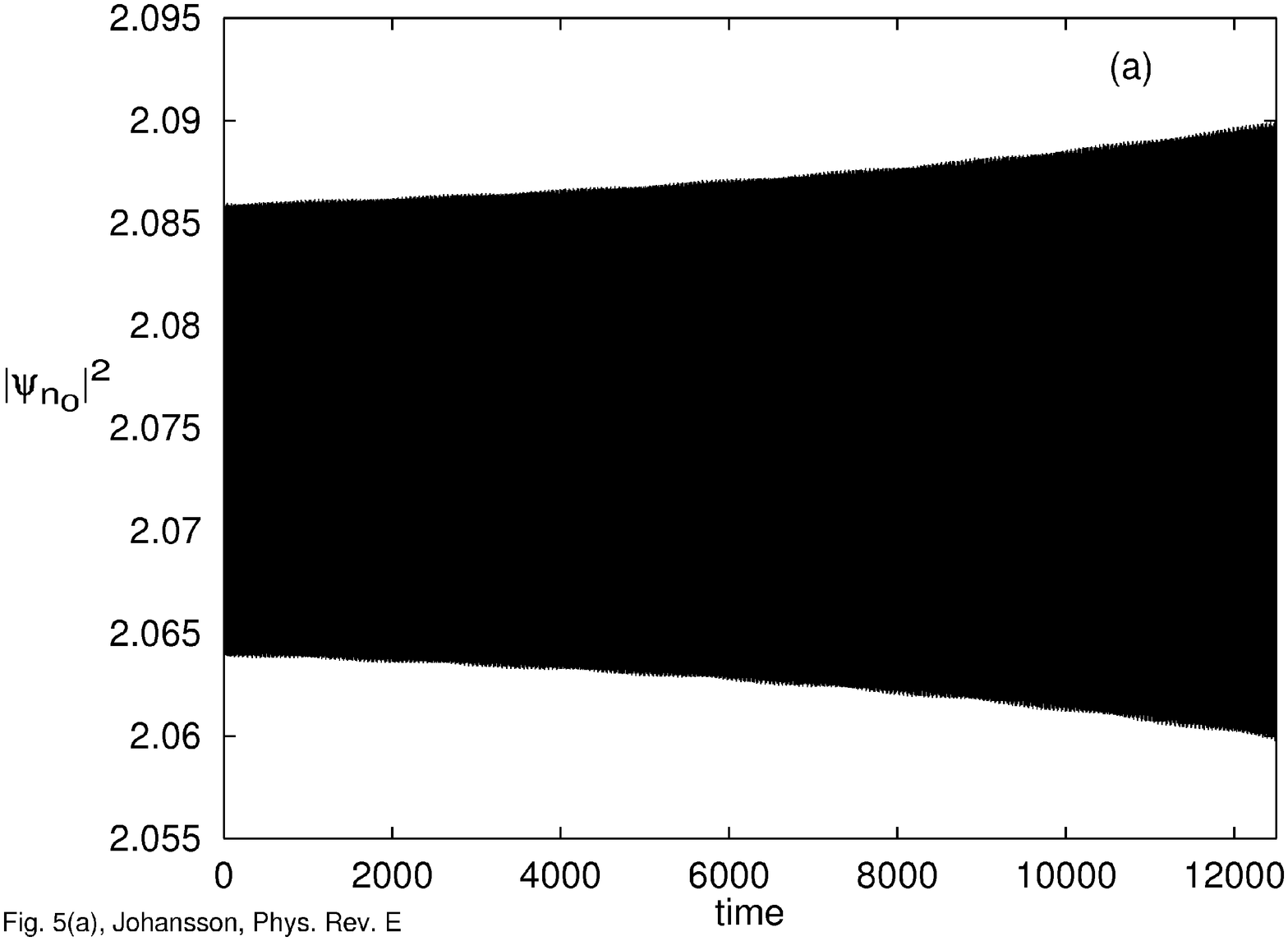}
\includegraphics[height=8.0cm,angle=270]{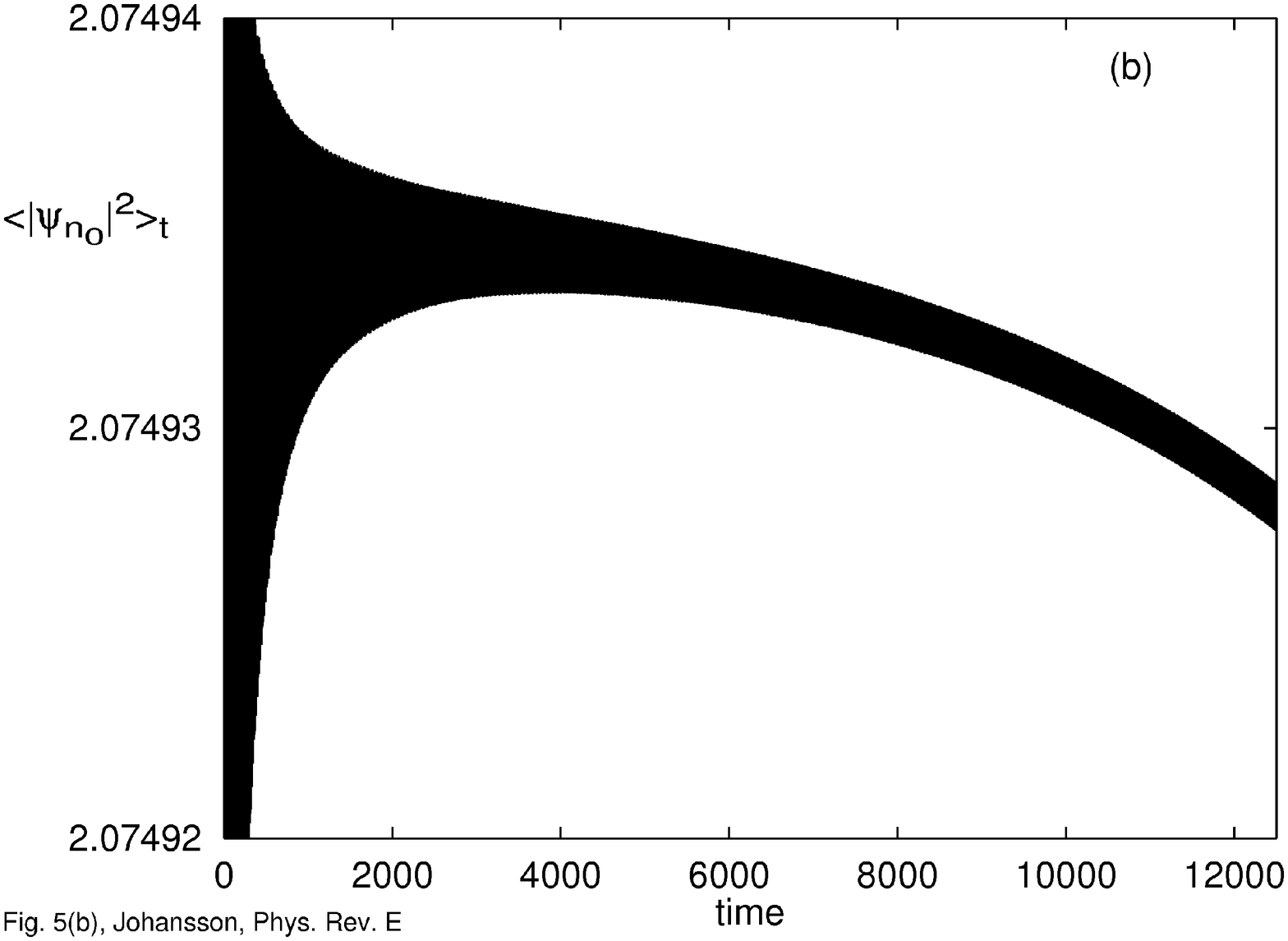}
\includegraphics[height=8.0cm,angle=270]{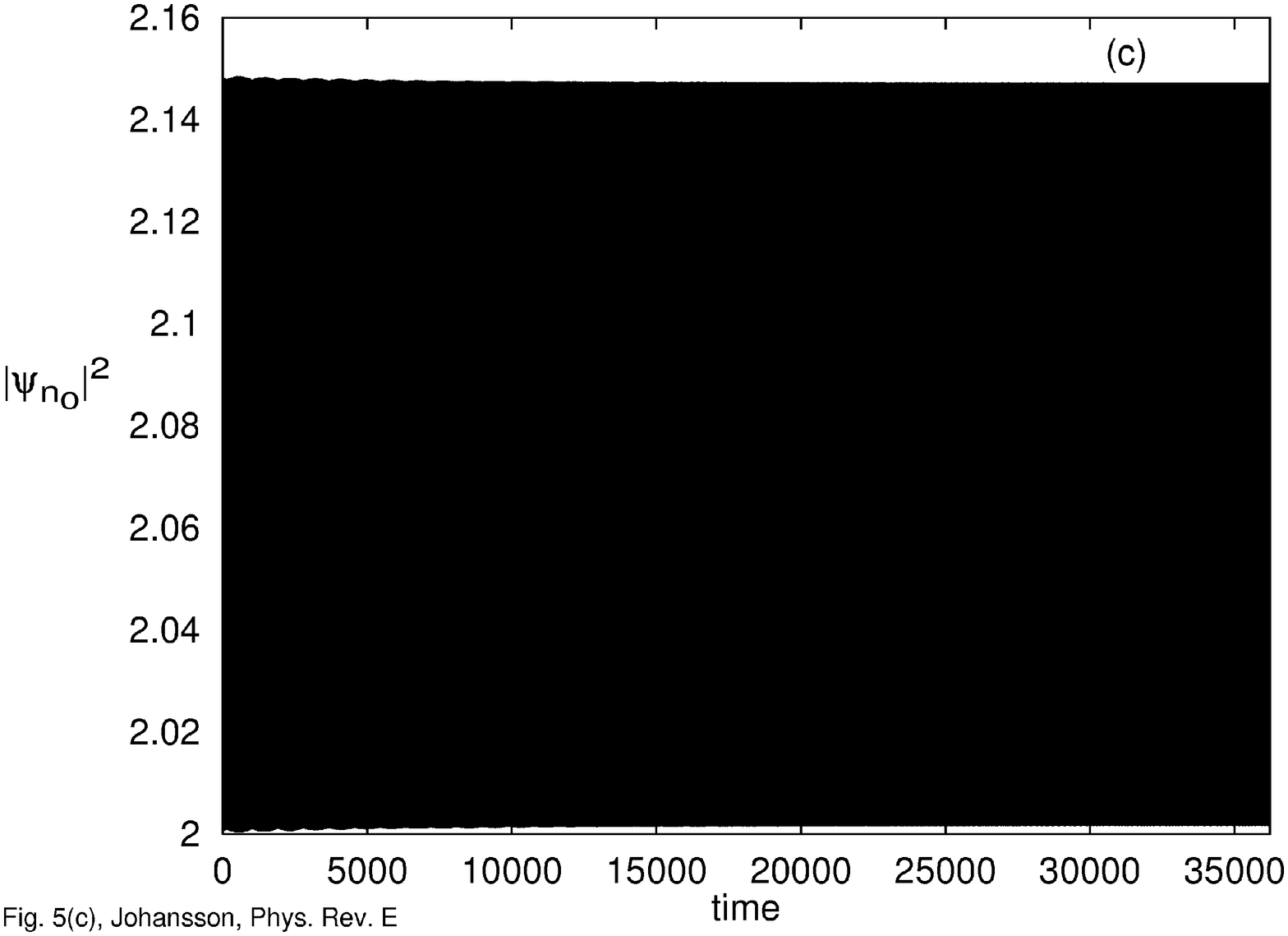}
\includegraphics[height=8.0cm,angle=270]{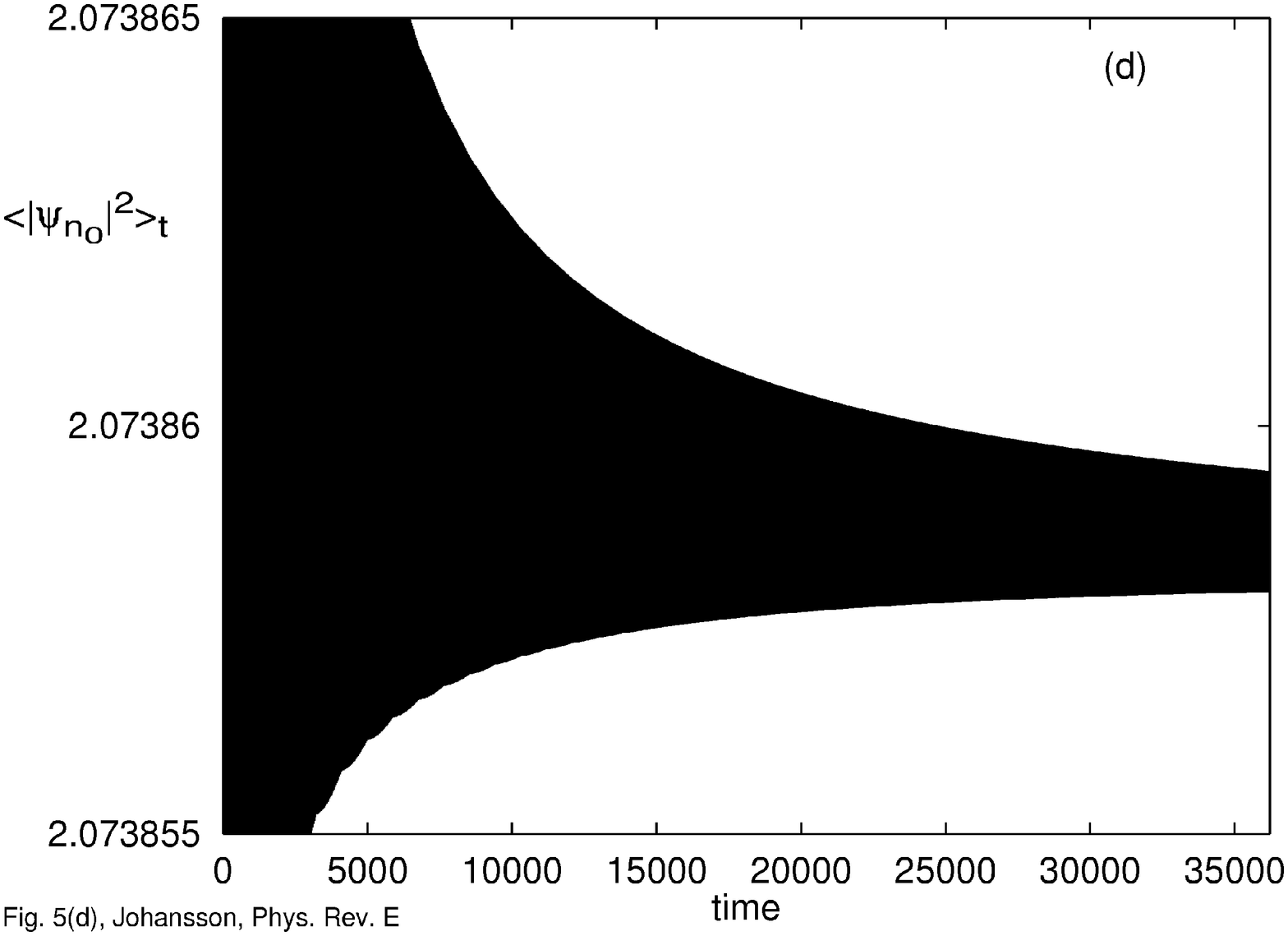}
\vspace{0.5cm}
\caption{Time-evolution of a breather with initial perturbations in the 
direction of extended eigenmodes with $q>q_c$. 
Parameters are $\Lambda=1.0$, $C=1$, and in (a), (b)
$\omega_p \approx 4.93$ ($q \approx 2.88$) and $a \approx 0.0644$, 
resp. in (c), (d) $\omega_p \approx 2.72$ ($q \approx 1.43$) and 
$a \approx 0.0383$. 
(a), (c) show the 
time-evolution of the central-site intensity $|\psi_{n_0}|^2$, while 
(b), (d) show its time-average.}
\label{fig5}
\end{figure}

Considering now the regime $q>q_c$ where all multiples of $\omega_p$ are 
outside the phonon band, the most important conclusion from our extensive 
numerical investigations is that we {\em never observe breather growth}. 
Instead, we sometimes (but not always) observe a very slow decrease of 
$\left \langle |\psi_{n_0}|^2 \right \rangle_t$, and an increase of the 
fluctuations around this mean-value. This behaviour is illustrated by 
Fig.~\ref{fig5} (a) and (b). A possible interpretation of these results is 
that, since the higher harmonics which are created by the breather-phonon 
interaction cannot propagate, they stay trapped around the breather. Thus, 
this could lead to  a transfer of energy from the 'pure' breather, which 
acquires more and more internal frequencies and becomes a 'chaotic breather' 
\cite{ThierryFPU}. Another possible interpretation is that the increase 
of the oscillation amplitude is connected with the oscillatory instabilities 
of the standing waves; as we will show in the next section these instabilities 
provide a mechanism for breather decay. However in some cases, illustrated 
by Fig.~\ref{fig5} (c) and (d), the oscillation amplitude  as well as its 
average value apparently approaches a constant limit value. We have at present 
no explanation for this behaviour (as will be discussed in the next section, 
there exist exact 'phonobreather' solutions which could be candidates for 
such  a final state, but they are unstable); it is possible that the time 
ranges that we were able to study with sufficient numerical accuracy in these 
cases simply were too short to observe the scenario described by 
Fig.~\ref{fig5} (a) and (b). 

To conclude this section, we repeat our main result that breather growth is 
observed if and only if $q<q_c$, where $q_c$ is given by (\ref{qc}). The 
fact that $q_c=0$ for $\Lambda>4C$ thus implies the existence of an upper 
limit beyond which the breather cannot grow with the type of perturbations 
considered here. We would also like to relate our results to recent numerical 
simulations of breathers interacting with {\em propagating} phonons in 
Klein-Gordon \cite{ThierryPhD} and FPU \cite{ThierryFPU} lattices. For the 
Klein-Gordon lattice with a (soft) Morse on-site potential, phonons with small 
wave vector 
$q$ were observed to yield breather growth, while phonons with large $q$ 
caused breather decay. For a FPU lattice with hard anharmonicity, the opposite
 situation was 
observed, i.e., small-$q$ phonons caused breather decay and large-$q$ phonons 
breather growth. The fact that the situation for the hard FPU lattice was 
 opposite 
to that of the soft Klein-Gordon lattice could be expected, since in the 
former case the modulational instability occurs for large $q$, whereas
soft Klein-Gordon and DNLS lattices with $C>0$ are modulationally 
unstable for small 
$q$. However, we stress that the relation between plane-wave modulational 
instability and breather growth is nontrivial, and at least for the case 
considered in this paper the critical value $q_c$ for breather growth from 
interaction with standing-wave phonons differs 
from the critical value $q=\pi/2$ for modulational instability of travelling 
waves.


\section{Breather growth and destruction from standing-wave instabilities}
\label{sec:instabilities}


In this section, with the aim at describing the interaction between a 
breather and a 
standing-wave phonon with non-negligible amplitude, we will take a slightly  
different point of view than in the preceding sections. Instead of  
choosing as initial condition an exact breather 
solution and adding a perturbation corresponding to an eigenmode of the 
linearized equations, we will here consider initial conditions which are exact 
{\em phonobreather} \cite{Aubry} (or {\em nanopteron}) solutions. By 
definition, a phonobreather consists of a spatially localized breather on top 
of a spatially extended tail which is a nonlinear, standing-wave phonon. 
(There are also solutions where the tail is a propagating wave 
\cite{AubryCretegny}, but 
we will not discuss them further here.) 
Phonobreathers exist 
generically for nonlinear lattice-equations (see e.g. \cite{Aubry,Marin96}), 
but their existence normally requires an integer relationship between the 
breather and phonon frequencies. However, for the DNLS equation 
phonobreathers exist for 
{\em any} (rational or irrational) relation between the two 
frequencies \cite{JohanssonAubry}, as a consequence of the additional 
invariance of the 
equation under global phase transformations.

Since the phonobreathers are exact solutions consisting of a breather part and 
a standing-wave part, one could expect them to be 
attractors for the initial conditions considered in Sec. \ref{sec:phonons}. 
However, as was shown recently \cite{Anna}, generically for soft Klein-Gordon 
and DNLS models with $C>0$ {\em all} phonobreathers with 
phonon wave vector $q \neq \pi$ will be {\em linearly unstable} if, for 
fixed phonon amplitude $a$, the linear coupling $C$ is larger than some 
threshold value $C_{cr}(a,q)$ (i.e., away from the anticontinuous limit). 
(For lattices with hard  potentials and DNLS with $C<0$, the stable 
phonobreather has $q=0$.)
These instabilities are caused by an 
oscillatory instability of the standing-wave phonon itself, which can be 
understood by considering the construction of a nonlinear standing wave 
with wave vector $q$ close to $\pi$ at the 
anticontinuous limit $C=0$ by introducing a periodic array of 
discommensurations or 'defects' in the nonlinear phonon with wave vector 
$\pi$ and amplitude $a$, 
$\psi_n=a(-1)^n e^{-i(4C-|a|^2)t}$, which is linearly stable for all $a$ and 
$C>0$
\cite{KivsharPeyrard}. In the anticontinuous limit $C=0$, each defect consists 
when $\pi/2<q<\pi$
of one extra site with $\psi_n=0$ added to the $\pi$-phonon, which 
consequently suffers an additional phase shift of $\pi$ across each defect. 
For $0<q<\pi/2$, each defect consists of several consecutive zero-amplitude 
sites with associated phase shifts (a general method for generating the 
anticontinuous coding sequence for standing-wave phonons from a circle map is 
described  
in  \cite{Anna}); in this case it is also useful to consider the periodic 
repetition of sites with $\psi_n=\pm a$ as defects of the zero-amplitude state.

The limit case of one isolated zero-amplitude defect, which is a discrete 
counterpart 
of the dark-soliton solution of the continuum NLS equation,  was investigated 
in \cite{JohanssonKivshar}. The linear stability analysis of this mode showed 
that, although it is stable close to the anticontinuous limit, it suffers a 
bifurcation for $C/|a|^2 =C_c\approx 0.0765$ where two pairs of eigenvalues of 
the eigenvalue problem (\ref{matrix}) go out in the complex plane. The 
resulting oscillatory instability occurs due to a resonance between a mode 
localized around the defect (the defect pinning mode) and linear radiation 
modes. It was shown that for finite systems, the mode recovers its stability 
above some upper critical value of $C/|a|^2$ (since the wavelength of the 
resonating linear modes becomes larger than the system size); 
however this critical value 
increases with system size so that in the limit of an infinite system, the 
instability persists for all $C/|a|^2 > C_c$ but with a growth-rate that 
decreases in an exponential-like fashion when approaching the continuum limit 
$C/|a|^2 \rightarrow \infty$. This instability was shown to result in the 
defect becoming mobile  (in NLS terms, the stationary 'black' soliton with 
zero minimum intensity transforms into a moving 'grey' soliton with non-zero 
minimum intensity) and radiation being emitted. In terms of the phase 
dynamics, this describes a moving, slowly spreading phase kink. 

The instability scenario for the standing-wave phonons is basically the same 
as for the isolated defect, with the essential difference that the localized 
pinning modes associated with the individual, periodically repeated  defects 
now will form a continuous 'defect band'. In general, the 
Krein signature \cite{Aubry} of this defect band is opposite to that of the 
bands associated with the non-zero amplitude sites \cite{Anna} 
(for $0<q<\pi/2$ there are
generally several defect bands, but they will have the same Krein 
signature), and as a consequence 
resonances between the bands will occur if the linear coupling $C$ is large 
enough, giving rise to similar oscillatory instabilities as described above 
(details are given in \cite{Anna}). We remark that earlier analysis \cite{KHS} 
of standing waves in nonlinear lattices, based on a quasi-continuum 
approximation, did not reveal these instabilities since their origin is the 
discrete nature of the lattice.

Let us now return to the main objective of this section, namely to study the 
effect of the oscillatory standing-wave instabilities on the phonobreathers. 
We find that the families of phonobreathers which are stable close to the 
anticontinuous limit and whose tails approach harmonic standing waves in the 
small-amplitude limit can be constructed from anticontinuous standing-wave
solutions at $C=0$, placing the breather at a zero-amplitude site 
of the phonon and adjusting it so that the resulting solution is either 
symmetric 
or antisymmetric around the breather site. Denoting the anticontinuous 
breather amplitude by 
$b$ (the phase of the breather site is unimportant when $|b|\neq |a|$ 
\cite{JohanssonAubry}), this yields the following 
possibilities:

\noindent (i) For $q>\pi/2$ the antisymmetric anticontinuous solution 
(here  $q=2\pi/3$)
\begin{equation}
 \{\psi_n (0) \} = \{...-a,a,0,-a,a,0,-a,a,b,-a,a,0,-a,a,0...\} ,
\label{i}
\end{equation}
with the asymptotic behaviour $\psi_n(0)\sim a \sin (qn), n\rightarrow \pm 
\infty$ in the continuum limit $C/|a|^2 \rightarrow \infty$. (Note that this 
solution is antisymmetric only at $C=0$.)

\noindent (ii) From (\ref{i}) we can construct a symmetric solution for 
$q>\pi/2$ by 
introducing an additional phase shift of $\pi$ at one side of the  breather 
site, giving
for $q=2\pi/3$
\begin{equation}
 \{\psi_n (0) \} = \{...-a,a,0,-a,a,0,-a,a,b,a,-a,0,a,-a,0...\} ,
\label{ii}
\end{equation}
with the asymptotic behaviour $\psi_n(0)\sim a \cos (qn \pm \pi/2), 
n\rightarrow \pm 
\infty$ in the continuum limit.

\noindent (iii) For $q=M\pi/N<\pi/2$, $N$ even, the number of consecutive 
zero-amplitude sites is odd and the antisymmetric anticontinuous solution is 
for e.g. $q=\pi/4$
\begin{equation}
 \{\psi_n (0) \} = \{...0,0,0,a,0,0,0,-a,0,0,0,a,0,b,0,-a,0,0,0,a,0,0,0,-a...\}
\label{iii}
\end{equation}
behaving as  $\psi_n(0)\sim a \cos (qn), n\rightarrow \pm 
\infty$ in the continuum limit.

\noindent (iv) From (\ref{iii}) the symmetric solution for  $q=M\pi/N<\pi/2$, 
$N$ even, is constructed by a phase shift as above, giving
for $q=\pi/4$
\begin{equation}
 \{\psi_n (0) \} = \{...0,0,0,a,0,0,0,-a,0,0,0,a,0,b,0,a,0,0,0,-a,0,0,0,a...\}
\label{iv}
\end{equation}
with the asymptotic behaviour $\psi_n(0)\sim -a \sin (qn \pm \pi/2), 
n\rightarrow \pm 
\infty$ in the continuum limit.

\noindent (v) For $q=M\pi/N<\pi/2$, $N$ odd, the number of consecutive 
zero-amplitude sites is even and we must add an extra site to obtain the 
antisymmetric anticontinuous solution, which 
for e.g. $q=\pi/3$ becomes
\begin{equation}
 \{\psi_n (0) \} = \{...0,0,a,0,0,-a,0,0,a,0,b,0,-a,0,0,a,0,0,-a,0,0,a...\}
\label{v}
\end{equation}
behaving as  $\psi_n(0)\sim a \cos (q(n+1/2)\pm q/2), n\rightarrow \pm 
\infty$ in the continuum limit. (A solution with similar properties  
is obtained by instead removing one zero-amplitude site; however its 
symmetric counterpart is always unstable.)

\noindent (vi) The symmetric counterpart of (\ref{v}) for  $q=M\pi/N<\pi/2$, 
$N$ odd, is constructed by a phase shift as above, giving
for $q=\pi/3$
\begin{equation}
 \{\psi_n (0) \} = \{...0,0,a,0,0,-a,0,0,a,0,b,0,a,0,0,-a,0,0,a,0,0,-a...\}
\label{vi}
\end{equation}
with the asymptotic behaviour $\psi_n(0)\sim -a \sin (q(n+1/2) \pm (q+\pi)/2), 
n\rightarrow \pm 
\infty$ in the continuum limit.

\begin{figure}[ht]
\noindent
\includegraphics[height=12.6cm,angle=270]{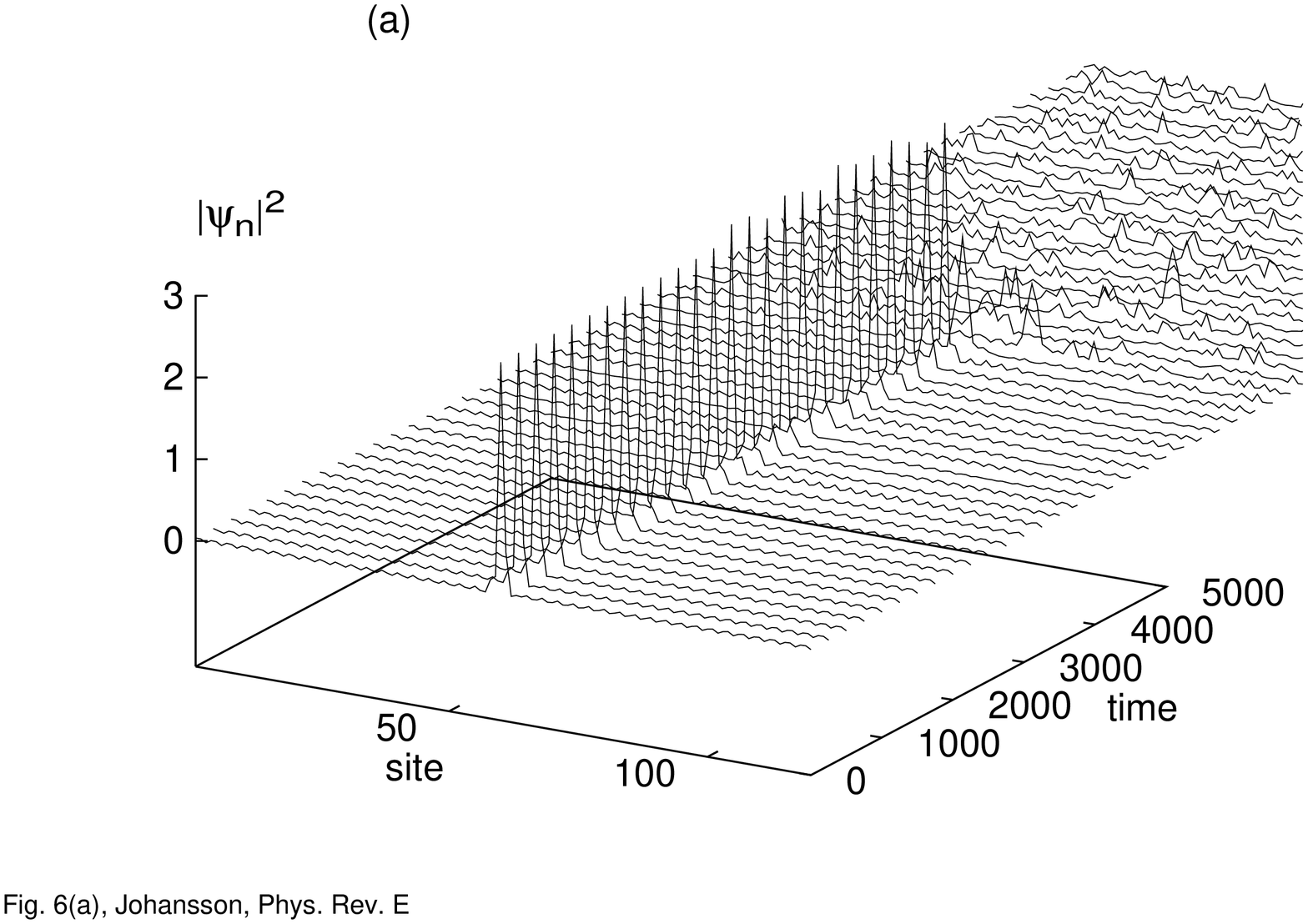}
\includegraphics[height=12.6cm,angle=270]{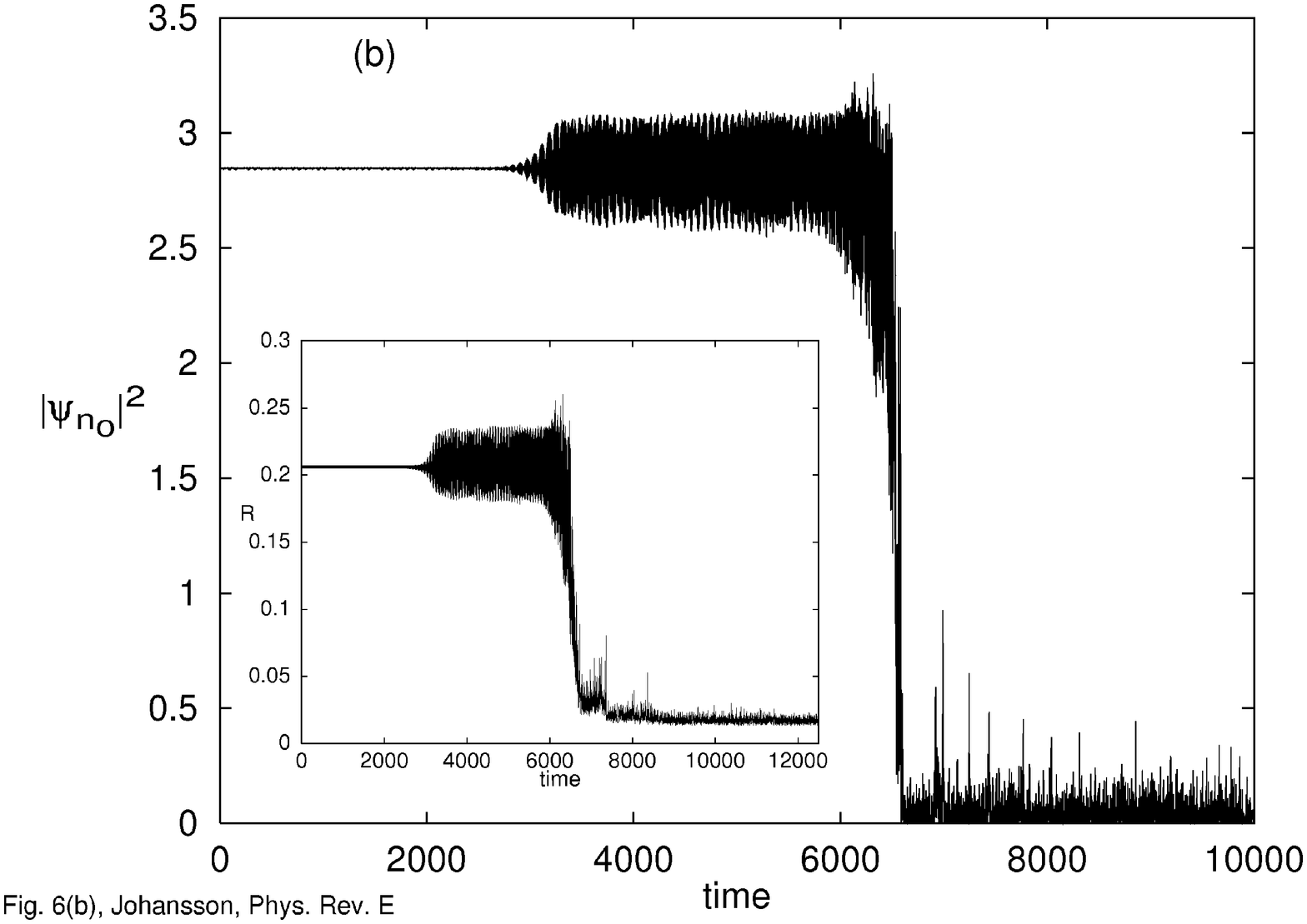}
\vspace{0.5cm}
\caption{Time-evolution of the phonobreather (\ref{i}) with $q=2\pi/3$ for a 
system of 120 sites (periodic boundary conditions), 
perturbed only by the numerical truncation errors. The breather 
frequency is $\Lambda_b=1.55$, phonon frequency $\Lambda_{ph}=-2.95$ (phonon 
amplitude $a\approx 0.2$), and $C=1$. (a) shows $|\psi_{n}(t)|^2$, while (b) 
(main figure)  
shows the intensity of the breather central site $|\psi_{n_0}|^2$. Inset in 
(b) shows the inverse participation number 
$R={\cal N}^{-2} \sum_n |\psi_n|^4$, which gives a qualitative measure of the 
degree of localization.}
\label{fig6}
\end{figure}

A typical example on the time-evolution for an initially very weakly perturbed 
phonobreather with phonon wave vector $q>\pi/2$ and phonon amplitude small but 
non-negligible compared to the breather amplitude is illustrated in 
Fig.~\ref{fig6}. (The example in the figure belongs to type (i), but similar 
dynamics is observed also for the spatially symmetric states of type (ii).) 
We can clearly distinguish two different steps leading to the final breather 
destruction. The first step is the linear oscillatory instability described 
above, which leads to the generation of new internal frequencies of the 
breather, and to the movement of the defect sites in a similar way as for the 
case of an isolated defect. In the second step, the moving defects start 
interacting, and a close inspection of Fig.~\ref{fig6} (a) shows that 
neighboring defects  tend to merge and create regions of accumulated phase 
fluctuations travelling around in the lattice. These will interact with the 
breather, and apparently cause its decay. When the breather has decayed 
sufficiently to have an excitable pinning mode, it will start to move in the 
lattice but with rapidly decreasing amplitude, and it will finally be 
destroyed. 
We have at present 
no complete understanding for the mechanism by which the interaction 
of the breather with the moving 
'phase kinks' cause its decay, but we remark that 
a similar scenario was observed when adding to
 the DNLS equation an external, parametric white noise term \cite{CGJK}. 
In that case, the white-noise approximation allowed a qualitative 
understanding of the breather decay as a consequence of phase fluctuations 
by using a collective coordinate approach.

However, to observe this scenario for  breather destruction it is necessary 
(at least for a finite system) 
that the phonon amplitude is not too small compared to the breather amplitude. 
If we increase the breather amplitude in Fig.~\ref{fig6} (or decrease the 
phonon amplitude) sufficiently, we find that although the oscillatory 
instability develops, the fluctuations created in the second step will be 
too weak to cause the breather to decay, and it will live seemingly forever as 
a 'chaotic phonobreather'. The absence of decay for small perturbations can 
be viewed as  a 
consequence of the fact that the single-site DNLS breather is nonlinearly 
(Lyapunov) stable for norm-conserving 
perturbations, in the sense that $|\psi_n(t)|$ remains arbitrarily close to 
the breather for all times if the initial perturbation is small enough 
\cite{Weinstein}. Thus, it is clear that for finite systems, the breather 
cannot be destroyed unless the phonon amplitude exceeds some critical value, 
while nothing can be said about the infinite system since any infinitely 
extended phonon obviously has an infinite norm.

\begin{figure}[ht]
\noindent
\includegraphics[height=13.75cm,angle=270]{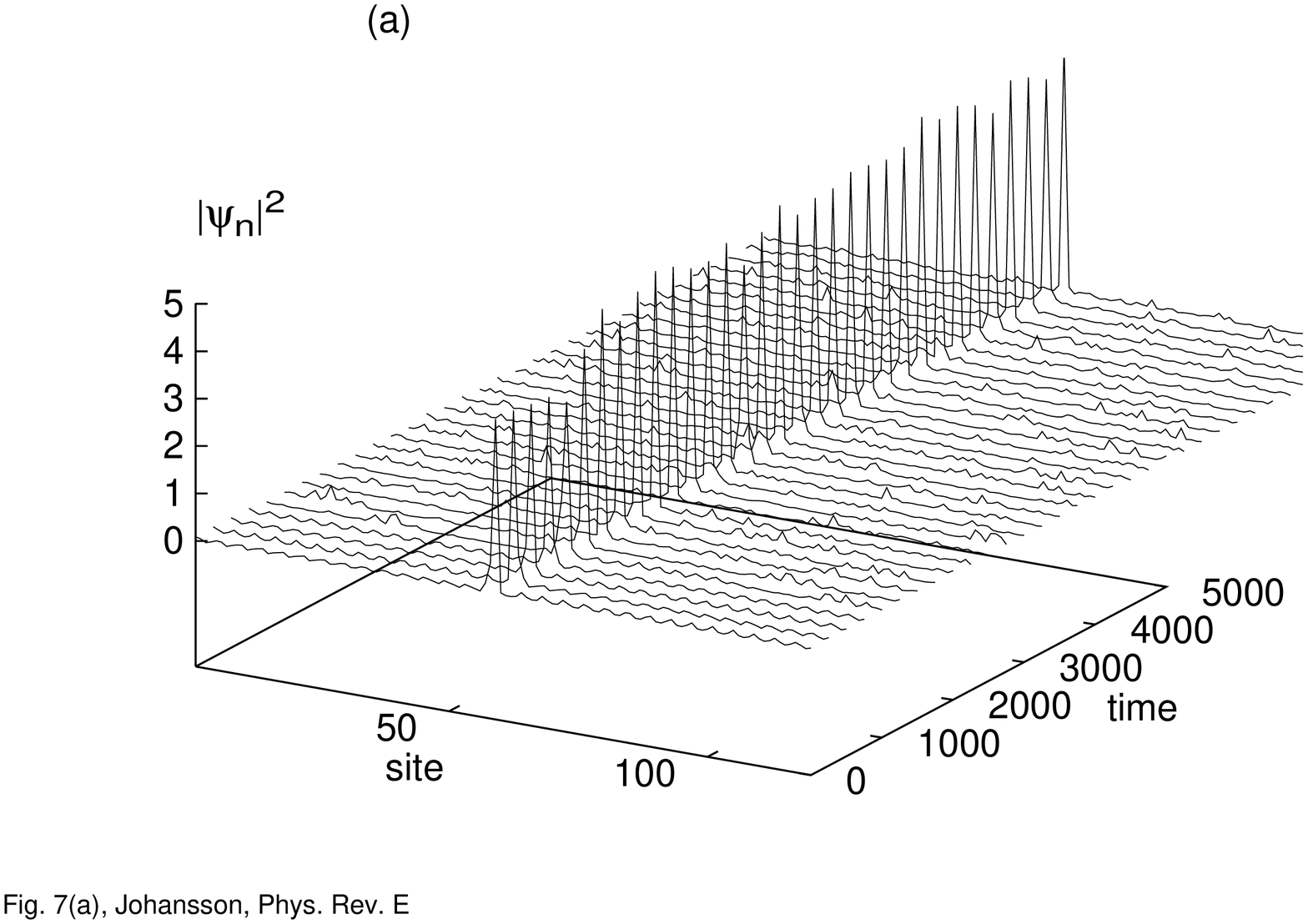}
\includegraphics[height=13.75cm,angle=270]{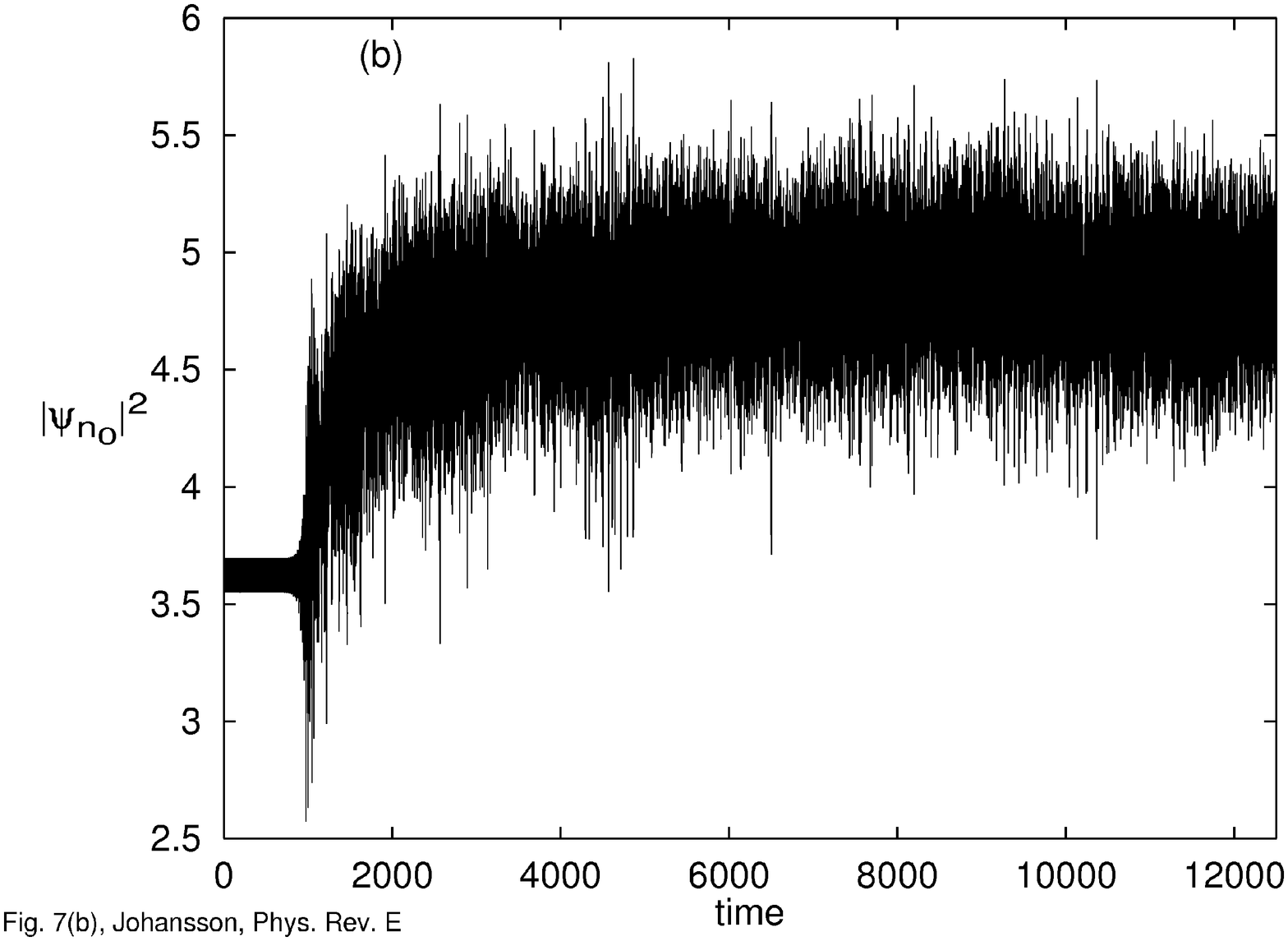}
\vspace{0.5cm}
\caption{Time-evolution of the phonobreather (\ref{iii}) with $q=\pi/4$ for a 
system of 120 sites (periodic boundary conditions), 
perturbed only by the numerical truncation errors. The breather 
frequency is $\Lambda_b=2.2$, phonon frequency $\Lambda_{ph}=-0.5$ (phonon 
amplitude $a\approx 0.3$), and $C=1$. (a) shows $|\psi_{n}(t)|^2$, while (b) 
shows the intensity of the breather central site $|\psi_{n_0}|^2$.}
\label{fig7}
\end{figure}

With $0<q<\pi/2$, the first step resulting from the oscillatory instability 
occurs in a similar way as for $q>\pi/2$: the breather 
acquires new frequencies and the small-amplitude sites of the phonon start 
moving. By instead making the interpretation that the sites with non-zero 
amplitude at the anticontinuous limit are defects in the zero-amplitude state, 
their movement can be seen as a consequence of the repulsive interaction 
between spatially separated, small-amplitude breathers with opposite phases 
observed e.g. in \cite{Aceves}. We also observe, similarly as for $q>\pi/2$, 
the merging of neighboring defects, but in this case their interaction with 
the breather will not lead to breather decay, but rather to breather growth 
if the original amplitude of the breather is not too large. A typical example 
is illustrated in Fig.~\ref{fig7}. It can be seen by a careful inspection of 
Fig.~\ref{fig7} (a) how the merging of small-amplitude sites results in 
localized humps of larger amplitude reminiscent of small-amplitude moving 
breathers travelling around in the lattice. The interaction of these humps 
with the original breather leads to growth of the latter in a similar way as 
observed in \cite{Kim,PeyrardKG,ThierryFPU}. However, this growth stops when 
the breather amplitude has reached a critical value which is close to (but 
apparently smaller than) that corresponding to the limit value $\Lambda=4C$ 
for small-amplitude perturbations found in Sec.~\ref{sec:phonons} (the latter 
corresponds to $|\psi_{n_0}|^2 \approx 5.65 $). The final state appears also 
here to be a 'chaotic phonobreather'; we have followed the time evolution of 
this kind of state for times up to $10^6$ without seeing any signs of decay. 
Also, if the initial breather frequency is chosen above the critical value 
$\Lambda=4C$, we typically do not observe breather growth; instead the mean 
value of the chaotic amplitude oscillations resulting from the oscillatory 
instability remains close to the initial amplitude.


\section{Concluding remarks}
\label{sec:concluding}


Investigating the interaction between discrete nonlinear Schr\"odinger 
breathers and small perturbations, we have found firstly that exciting an 
internal mode of the breather always leads to a slow energy transfer to the 
breather, i.e., to breather growth. Furthermore, we found that a DNLS-breather 
can pump energy from a small-amplitude standing-wave phonon, provided that 
the phonon wave vector is smaller than the critical value $q_c$ given by 
Eq.~(\ref{qc}). In both cases, the mechanism for breather growth involves the 
higher-order generation of radiating modes. Since this mechanism disappears 
at the threshold value $\Lambda=4C$ of the breather frequency, it is 
impossible for a breather to grow beyond this value with the type of 
small-amplitude perturbations considered here. To analyze the interaction 
between breathers and standing-wave phonons of small but non-negligible 
amplitude, we considered the long-time evolution of weakly perturbed exact 
phonobreather solutions. The instabilities of these, originating in 
oscillatory instabilities of the nonlinear phonons, where shown to lead to 
propagating 
inhomogeneities whose interaction with the breather provided a mechanism for 
breather decay and destruction (when the phonon wave vector $q>\pi/2$) or 
growth ($q<\pi/2$ and $\Lambda < 4C$). 

As was mentioned already in the introduction, the existence of the two 
conserved quantities (\ref{DNLSham}) and (\ref{norm}) makes the DNLS equation 
non-generic among nonlinear lattice equations, and it is therefore necessary 
to investigate to what extent the results obtained in this paper apply also 
for Klein-Gordon and FPU lattices. We plan to address these questions in a 
forthcoming publication, but let us stress already here that the 
perturbational approach used here for the DNLS equation needs to be modified 
to 
account for the fact that generically, the dynamics of the breather involves 
also higher harmonics of its fundamental frequency. Moreover, the approach 
used in Secs.~\ref{subsec:conservation} and \ref{sec:phonons} based on the 
conservation laws cannot be directly applied in the absence of a second 
conserved quantity. However, in view of the wide applicability of the DNLS 
equation (and in particular its appearance as a limit case of general lattice 
equations as mentioned in the introduction), we believe that the mechanisms 
for breather growth and destruction described in this paper are essential 
ingredients also for the corresponding processes in general lattice models.


\acknowledgements

We thank Yu. S. Kivshar for giving us an early preprint of Ref.~\cite{PKA}, 
I. V. Barashenkov for directing our attention to Ref.~\cite{ABP}, 
and A. M. Morgante for discussions on phonobreathers and  standing-wave 
instabilities. 
M. J. acknowledges a Marie Curie
Research Training Grant from the European Community. A preliminary version 
of these results was presented at the conference Nonlinearity `99 
(Heraklion, May 10-14, 1999).


%
%

\newpage

\newpage

\newpage

\newpage

\end{document}